\documentclass[prd,superscriptaddress,onecolumn,showpacs,11pt,msmath,
preprintnumbers,showkeys]{revtex4}
\usepackage{wrapfig,rotating}
\usepackage{graphicx,epsfig,color}
\usepackage{subfig}
\makeatletter

\newenvironment{figurehere}
{\def\@captype{figure}}
{}
\makeatother

\usepackage{graphicx}
\usepackage{dcolumn}
\usepackage{bm}

\def\beq{\begin{equation}}
\def\eeq{\end{equation}}
\def\beeq{\begin{eqnarray}}
\def\eeeq{\end{eqnarray}}

\def\cO#1{{\cal{O}}\left(#1\right)}

\def\2GPD{$_2\mbox{GPD}$}

\def\12{$1\otimes 2$}
\def\22{$2 \otimes 2$}

\def\Qsep{Q_{\mbox{\rm\scriptsize sep}}}
\def\Qsep2{Q^2_{\mbox{\rm\scriptsize sep}}}

\begin{document}

\title{Double parton interactions  initiated by  direct photons in $\gamma p$ and $\gamma A$ collisions revisited.}
\author{B.\ Blok, R. Segev,
\\[2mm] \normalsize  Department of Physics, Technion -- Israel Institute of Technology,
Haifa, Israel}

\begin{abstract}We consider the double parton scattering (DPS) initiated by direct photons 
in $\gamma p$ and $\gamma A$ collisions. We extend the previously known results for photoproduction
to the case of electroproduction with nonzero photon virtuality $Q^2$, and then consider the DPS for both photo and electroproduction 
for HERA and IEC ( electron ion collider) kinematics. The number of DPS events at IEC is shown to be of the same order
as in HERA and is even larger, with increase in luminocity compensating the decrease of  the center of mass energy.
 \end{abstract}
\maketitle

\thispagestyle{empty}

\vfill
\section{Introduction}
\par The study of double parton scattering (DPS ), and multiparton interactions (MPI) started in early eighties \cite{TreleaniPaver82,mufti}.In recent years there 
was a considerable progress in the subject,
both theoretically~\cite{stirling,BDFS1,Diehl,stirling1,BDFS2,Diehl2,BDFS3,BDFS4,Diehl:2017kgu,Manohar:2012jr,ST,BSW} , see also the recent review \cite{book}, and experimentally. The recent applications however dealt mostly with pp and pA collisions.
On the other hand,  the MPI, including DPS in deep-inelastic scattering attracted relatively little attention, except some research
on MPI in HERA, where it was claimed that the inclusion of MPI improves the agreement between experiment and theoretical 
description of HERA meausurments, although no concrete evidence for MPI in HERA was presented experimentally
\cite{Yung1,Yung2}. More recently,
MPI in DIS were studied in \cite{CR} and in \cite{B}, where the contribution of MPI in the underlying event 
in future electron-ion collider (IEC)  was considered.
The MPI and in particular DPS are important both from the point of view of searching for new 
physics as irreducible backgrounds, as well as for the study of the perturbative and nonperturbative correlations of partons in the nucleon and in the nuclei.
\begin{figurehere}
\begin{center}
\includegraphics[scale=0.3,angle=90]{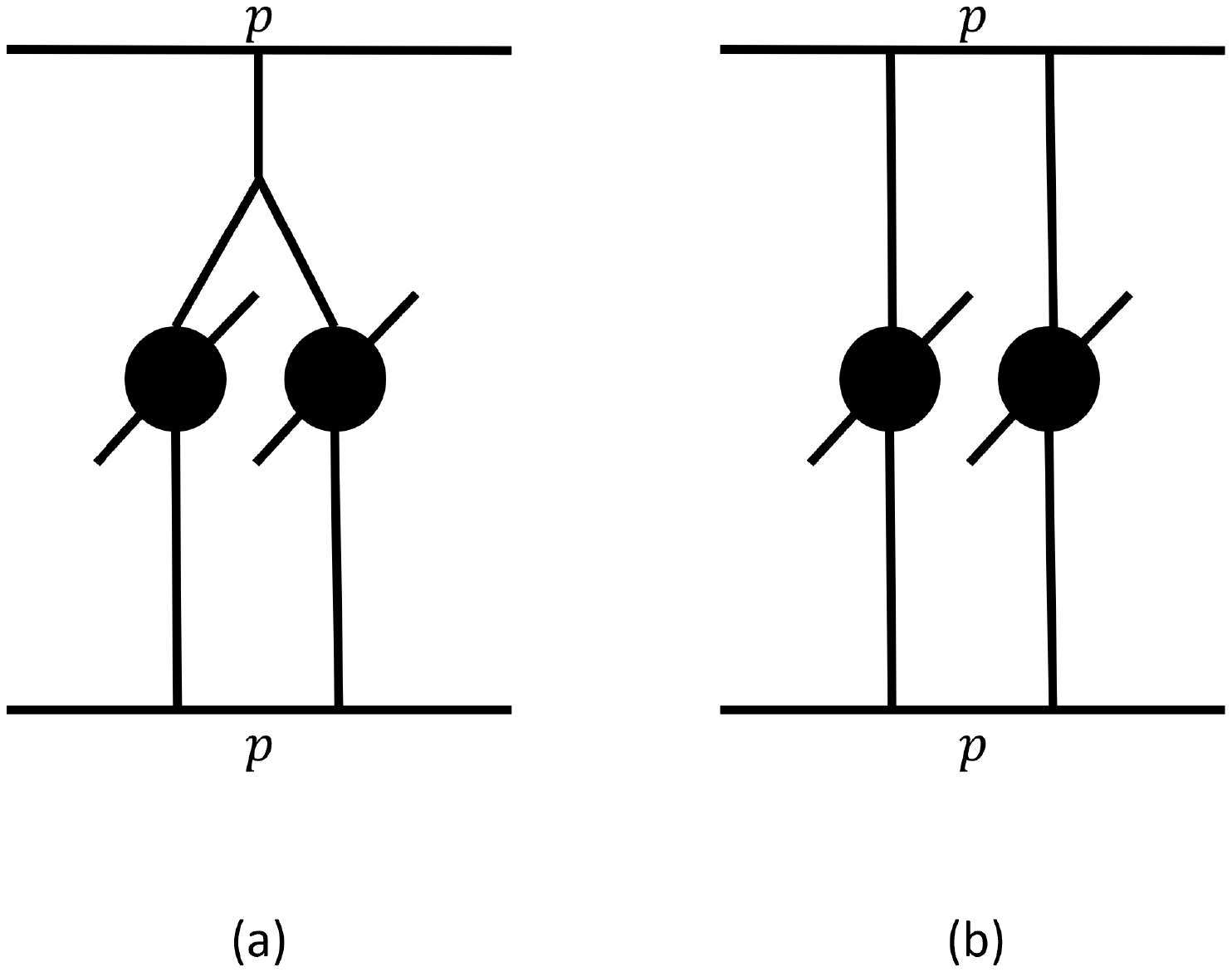}
\end{center}
\label{ran1}
\caption{the \12 and \22 mechanisms for DPS in pp collisions.}
\end{figurehere}
\par One of the new ideas suggested recently to explain the observed cross sections of DPS in pp and pA scattering was the so called \12 
mechanism. While in conventional DPS two independent partons from one nucleon interact with two independent partons from the second nucleon, in  \12 mechanism (see Fig. 1a) perturbative gluon from one of the nucleons involved in the collision splits into two and interacts with two independent partons of the second nucleon. The analysis in \cite{BDFS3,BDFS4}  shows that the contribution of this mechanism in pp and pA colllisions has the same order of magnitude as the contribution of conventional mechanism due to independent partons (see Fig.1b). However in pp and pA collisions it is not clear how to isolate this mechanism, and get the conclusive confirmation of its existence.
\par The photon nucleus and photon nucleon collisions give a unique possibility to separate such mechanism. Namely,
we can consider the so called direct photons,  The wave function of such photons can be viewed as corresponding 
to perturbatively created quark-anti-quark pair, where the quarks can be either light or heavy, like charmed quark.
In distinction to deep-inelastic scattering we can consider the processes where each of the created quarks interacts 
with a parton from the target nucleon, creating a quark and gluonic jets, where the gluon comes from the nucleon side.
Such processes can be especially easily identified if one of the quarks is a charmed quark, and we can neglect the charm-anti charm  content of the nucleon. Such processes  are the direct analogues of the \12 mechanism, see Fig. 2.
\begin{figurehere}
\begin{center}
\includegraphics[scale=0.3,angle=90]{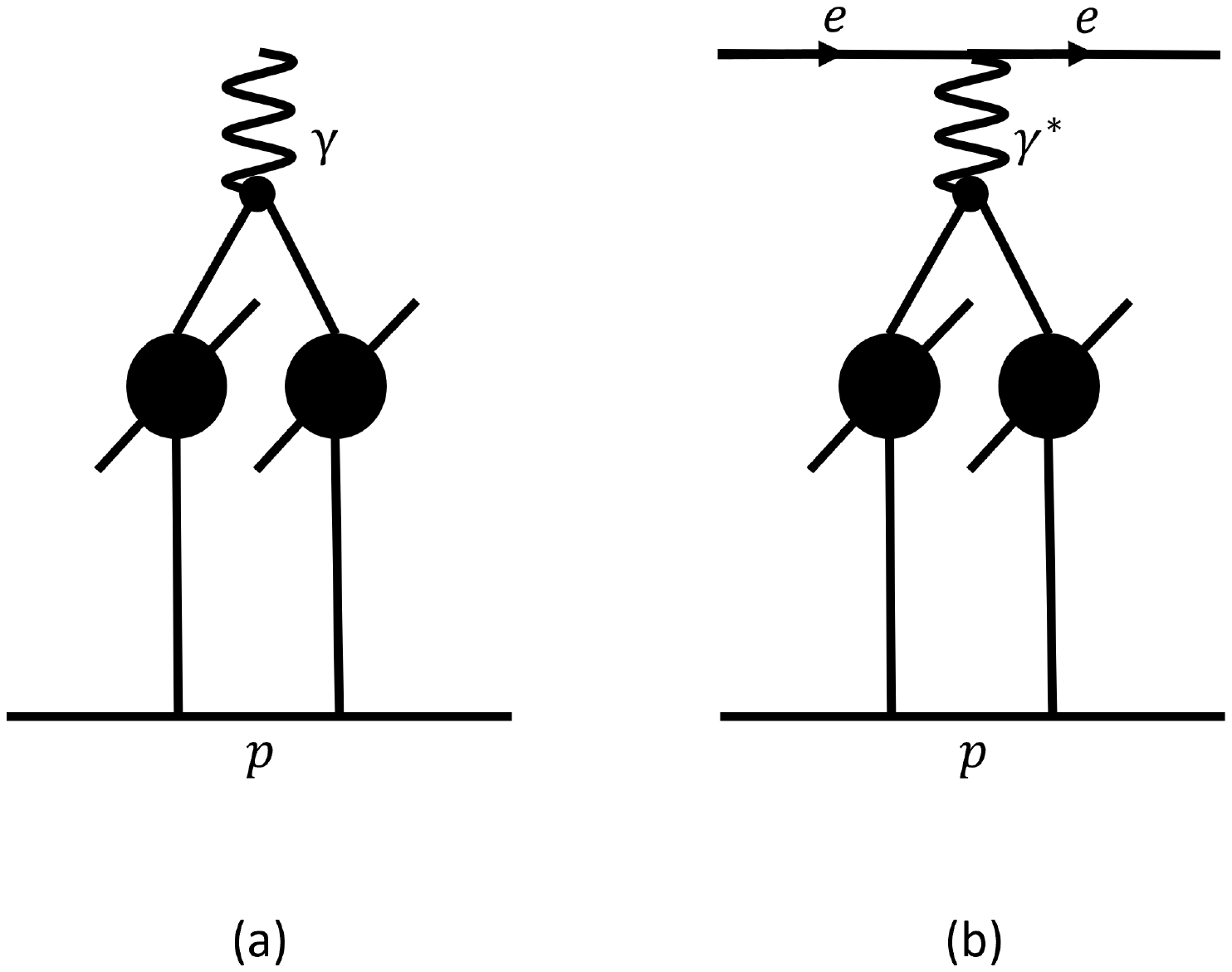}
\end{center}
\label{iren2}
\caption{The  DPS processes with the direct photons for  $\gamma p$ deep inelastic  collisions.}
\end{figurehere}
We expect such processes to be dominant in DPS at large $x_{ \gamma}\equiv x_1+x_2\sim 1$,
where $x_1,x_2$ are 
the Bjorken fractions of quark jets.
  In this paper we consider the case  when both $x_1,x_2\ge 0.2 $, are 
the Bjorken fractions of quark jets.
\par Such processes for the case of the photoproduction (see Fig.2a)
 were first considered in \cite{BS},
 where the DPS associated 
with direct photons were considered for HERA and LeHC kinematics and for ultraperipheral collisions at LHC.
 In this paper we
extend the results of \cite{BS} to the general case of electroproduction 
, i.e. we consider the case of virtual photons
$Q^2\ne 0$ (Fig.2b) .
In the limit $Q^2\rightarrow 0$ our results  correspond to photoproduction and we correct  several arithmetic mistakes made in numerical simulations in \cite{BS}. We consider the DPS due to direct photons both in the case of HERA and the future electron-ion collider (IEC).
Such processes may be an important part of the underlying event (UE)  in these colliders \cite{B}. 
We see that although the energies of HERA are 4 times larger than the maximal energy at IEC, this is compensated by much larger luminocity. We shall consider the maximum energy of $W=22500=150^2$ GeV$^2$,  where we shall assume luminocity as much as 100 times larger than in HERA,
and the energies $W=10000$  GeV$^2$where we assume luminocity 1000 times larger than in HERA.  Say,  for photoproduction with charmed jets we obtain for $p_t=3$ GeV approximately  $4* 10^4$  events per year  for total luminocity 
$\sim 1.7*10^{31} {\rm sm}^{-2}{\rm s}{-1}$ for HERA,
and of order $10^5$  events per year at IEC. The increase of the luminocity compensates the decrease in the center  of mass energy.

\par Our results may permit both to study short range correlations in a nucleon and nuclei and study the structure 
of underlying event at IEC\cite{B}. We shall consider two important final state cases. First is that each hard event has two light quark and two gluon jets, and the second case is two gluonic jets and two charmed quark states.
\par The paper is organized in the following way. In section 2 we calculate the  partonic model contributions to DPS cross section for transverse and virtiual 
photons (some technical details of the calculations are  given in Appendix A). In particular we show how to use gauge invariance to calculate the contribution due to longitudinal photons.
 In section 3 we calculate a total number of DPS events in different 
kinematics, both for photo and electroproduction, and include the pQCD evolution. In section 4 we reconsider numerical results  for photoproduction in HERA and compare them with the results for IEC. We also  consider the  numerical results  for electroproduction
for IEC  and HERA kinematics. 
Our results are summarized in section 5. In Appendix A we give the details of calculation of DPS cross -sections in Breit system. In appendix B we study how  the gauge invariance is realized.

\section{Basic formulae for MPI in the direct photon - proton scattering for given $Q^2$.}
\subsection{DPS kinematics.}
The DPS kinematics is similar to the kinematics of DIS process \cite{Frixione:1993yw}
\beq
\frac{d\sigma}{dQ^2dy}=\frac{\pi e^2}{4W}\frac{L_{\alpha\beta}D^{\alpha\nu}D^{\beta\mu} W_{\mu\nu}}{(2\pi)^3}
\eeq
where we  integrated over the azimuthal angle $\phi_e$ of the scattered electron.
Here $W=2kp$ is the center of mass energy squared  of the proton and electron, while $s=2qp$ is the centre of mass energy squared of the photon and proton,
$s=yW$, $q=k'-k$, where q is the momentum of the virtual photon, $q^2=-Q^2$,  $k',k$ are the final and initial momenta of the electrons that 
emit the virtual photon, and p is the momentum of the proton.
The propagator of the virtual photon $D_{\mu\nu}$ is conveniently  represented as 
\beq
D_{\mu\nu}(q)=-\frac{(g_{\mu\nu}-\frac{q^\mu q^\nu}{q^2})}{q^2}=-(e_L^\mu e_L^\nu-\sum_{\lambda=1,2}e_T^{\lambda\nu}e_T^{\lambda \nu})/q^2,
\eeq
where 
\beq
e_L=(q+2xp)/Q,\,\,Q\equiv \sqrt{Q^2}
\eeq
is the polarisation vector of the longitudinal gluon , and $e_T^\lambda, \lambda=1,2$ are the  basic polarisation vectors
 of the transverse photon,
The leptonic tensor $L_{\mu\nu}$ is given by 
\beq
L_{\mu\nu}=4(k^\mu k^{'\nu}+k^\nu k^{'\mu}-g^{\mu\nu}Q^2/2),
\eeq
 the tensor $W^{\mu\nu}$ is the hadronic  tensor which we shall start to calculate in the next subsection, and $x=Q^2/(2pq)$.

\subsection{Parton Model.}
\par  In this subsection we consider the 
the process of production of two dijets in the parton model. 
The corresponding kinematics is depicted in Fig. \ref{Fig3}, and is analogous to the  \12 transition in $pp$ collisions as it was 
mentioned in the Introduction.
Let us parametrize the momenta of quarks and gluons using Sudakov variables ($k_1,k_2$ are momenta of virtual  quarks and antiquark  of the  $q\bar q $ pair  and $k_3,k_4$ are the gluon momenta).
Let us analyze the lowest order  amplitude shown in Fig.3 for the  double hard collision which involves photon splitting.
\begin{center}
\begin{figurehere}
 \includegraphics[height=10.5cm,angle=90]{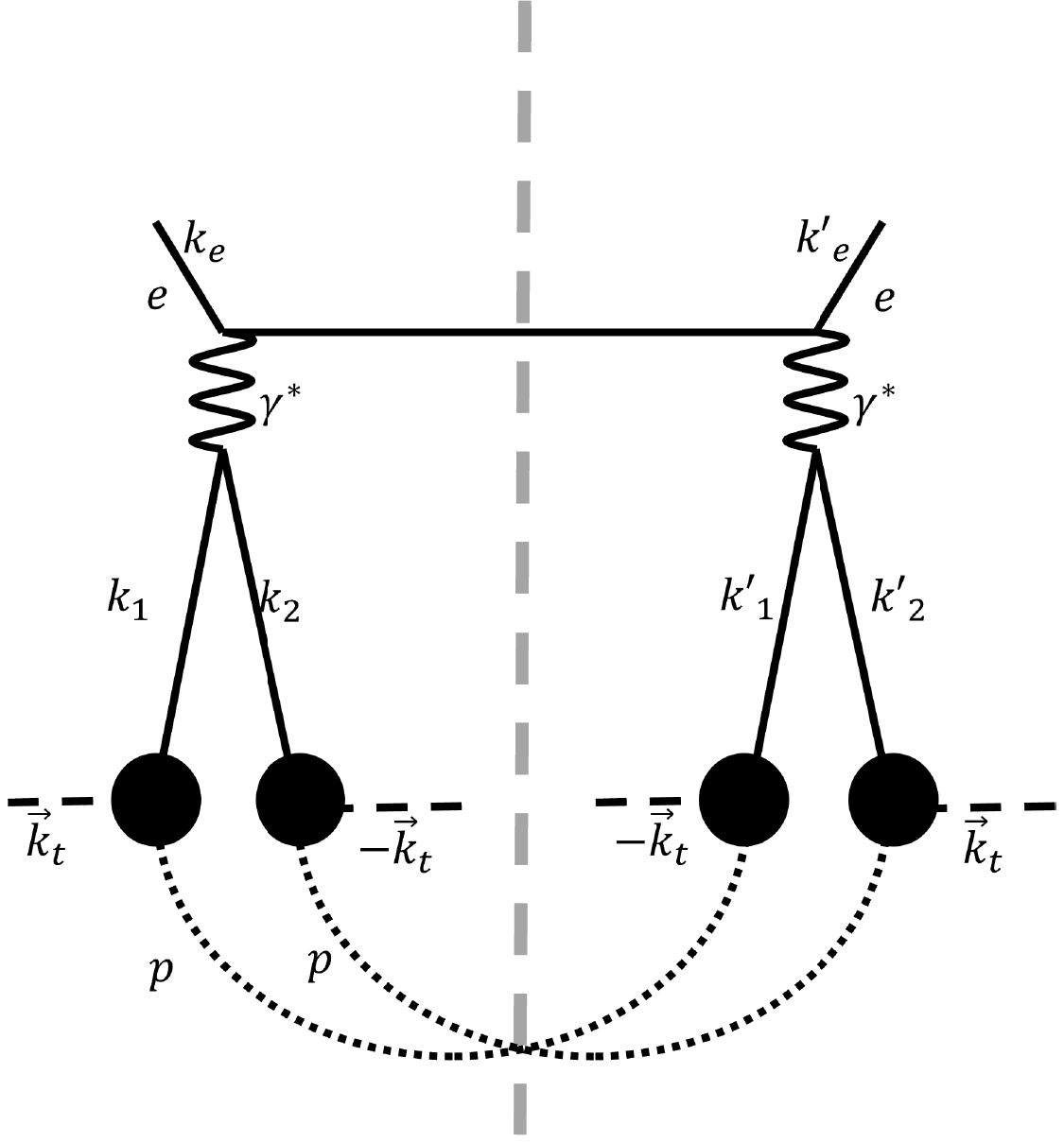}
\caption{\label{Fig3} parton model  for electroproduction with DPS.}
\end{figurehere}
\end{center}

We decompose parton momenta $k_i$ in terms of the Sudakov
variables using the light-like vectors $q'$ and  $p$ along the incident photon and proton momenta (here :
\begin{eqnarray}
k_1 &=& x_1q' + \beta p + k_t, \quad  k_3 \simeq (x_3-\beta)p; \nonumber \\
k_2 &=& x_2q' - (\beta+x) p - k_t, \quad k_4 \simeq (x_4+\beta -x)p; \nonumber \\
 \vec{k}_t &=& \vec{\delta}_{12} = -\vec{\delta}_{34} \> (\delta'\equiv0); \> k_0 \simeq(x_1+x_2)q. \nonumber
\end{eqnarray}
Here $q'=q+xp, q^{'2}=0$, and we neglect 
the nucleon mass: $p^2=0$ . Note that $x_1+x_2=1, q=k_1+k_2$ and $x=Q^2/(2pq)$.
For the case of massive, i.e. charmed quark, we shall neglect the charm quark masses except while dealing with infrared singularities and phase space constraints.
The light-cone fractions $x_i$, (i=1,..4), are determined by the jet kinematics (invariant masses and rapidities of the jet pairs).
The momenta $\vec \delta_{12},\vec \delta_{34},\delta'\equiv \delta_{12}+\delta_{34}$ are dijet  transverse imbalances, in the parton model their  sum is zero due to momentum conservation.
The fraction $\beta$ that measures the {\em difference}\/ of the  longitudinal momenta of the two partons coming from the hadron, is arbitrary and is integrated over.
The fixed values of the parton momentum fractions $x_3-\beta$ and $x_4+\beta$ correspond to the plane wave description of the scattering process in which the longitudinal distance between the two scatterings is arbitrary. This description does not correspond to the physical picture of the process we are discussing, where two partons originate from the same bound state.
In order to ensure that  partons $3$ and $4$ originate from the  {\em same hadron}\/ of a finite size, we have to introduce
integration over
$\beta$ in the amplitude, in the  region $\beta=\cO{1}$, as  was explained in detail in \cite{BDFS2}.

The Feynman amplitude contains the product of two virtual propagators. The virtualities $k_1^2$ and  $k_2^2$ in the denominators
of the propagators can be written  in terms of the Sudakov variables as
\beq
  k_1^2 = x_1\beta s -k_\perp^2, \>\>  k_2^2 = -x_2(\beta +x)s -k_t^2,
  \eeq
where $k_t^2\equiv (\vec{k}_t)^2>0$ the square of the two-dimensional transverse momentum vector.

The singular contribution we are looking for originates from the region $\beta\ll1$. Hence
the precise form of the longitudinal smearing
does not play role and the integral over $\beta$ yields the amplitude $A$
\beq
A\sim\!\! \int\! \frac{d\beta}{(x_1\beta s -k_t^2+\!i\epsilon)(-x_2(\beta+x) s -k_t^2+\!i\epsilon)}
=  \frac{1}{s(k_t^2+x_1x_2Q^2)}. \nonumber
\eeq
The numerator of the full amplitude is proportional to the {\em first power of the transverse momentum}\/ $k_t$.
As a result, the squared amplitude (and thus the differential cross section) acquires the necessary factor $1/k_t^2$
that enhances the back-to-back jet production.
\par The integration over $k_t$ gives a single log contribution to the cross section
 $\alpha_{\rm em}\log(Q_1^2/\mu^2)$ for light quarks and 
 $\alpha_{\rm em}\log(Q_1^2/m_c^2)$, where $Q_1\sim Q_2$ is the characteristic transverse scale of the hard  processes.
 The scale $\mu$ is the scale beyond which the nonperturbative dynamics becomes dominant
 and is of order of$m^2_\rho\sim 4m^2_{\rm const}$, where $m_\rho $ is the $\rho$- meson mass, $m_{\rm const}\sim 200-300$MeV  is the constituent quark mass. for light quark.

Note, that strictly speaking the answer due to diagram 3 is proportional to $\delta(\vec k_{1t}+\vec k_{2t})/(k_{1t}^2+x_1x_2Q^2)$, where $k_{1t}$ and $k_{2t}$ are the transverse momenta of quark and antiquark
( $\vec k_{1t}=-\vec k_{2t}\equiv \vec k_t$ in the parton model).

 The parton model answer is only single collinearly enhanced,
while we are looking for the double collinear enhanced contributions \cite{BDFS2}. It is well known that these contributions originate  from the gluon dressing of the parton model vertex,
with the $\delta$ function becoming a new pole.

\begin{figurehere}
\begin{center}
\includegraphics[scale=0.3,angle=90]{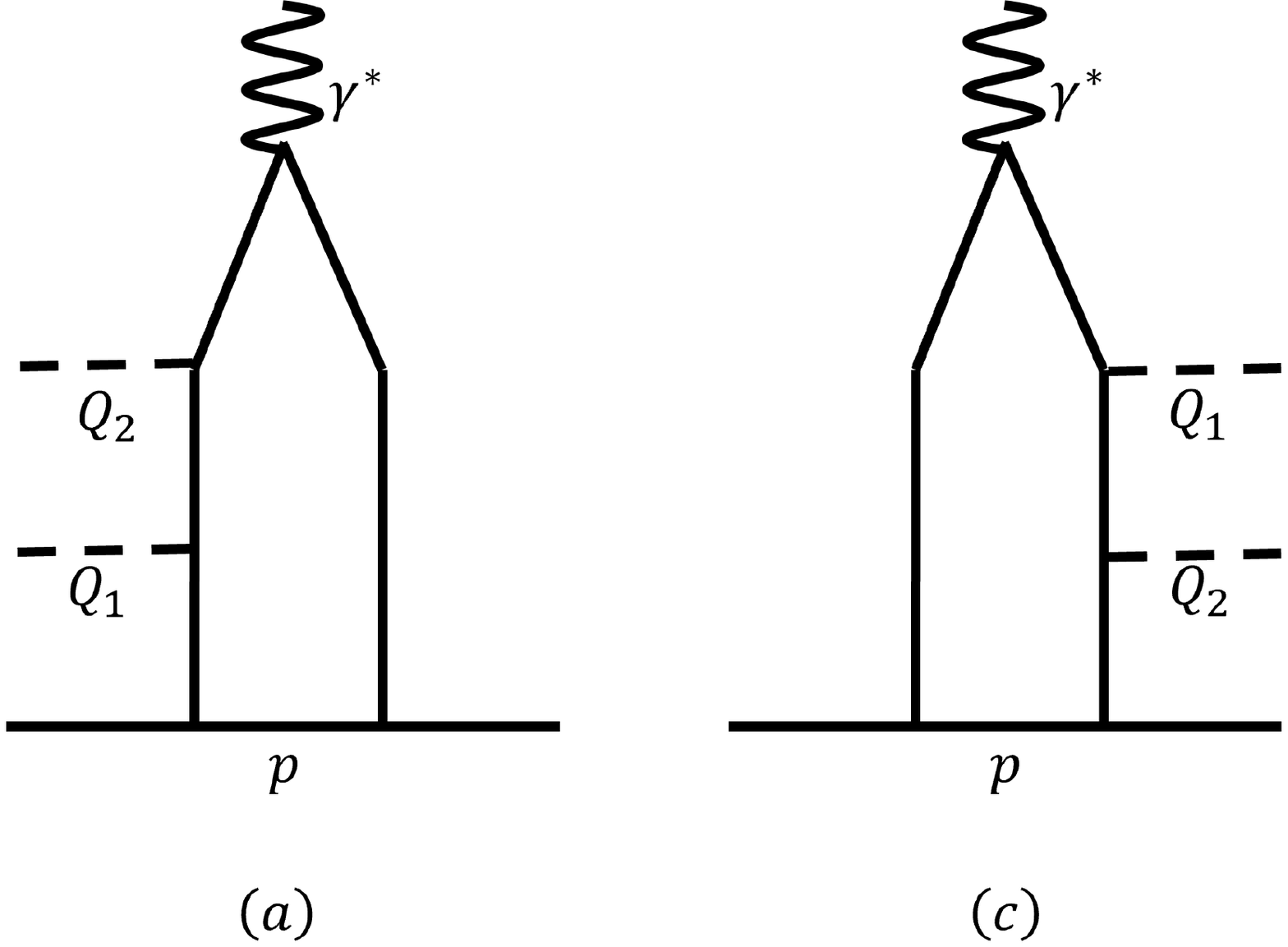}
\end{center}
\label{biga}
\caption{Gauge invariance and parton model.$Q_1$ and $Q_2$ are the dijet momenta.}
\end{figurehere}
\par Assuming factorisation theorem in order to calculate the vertex function it is enough to calculate the diagram 3.
The direct calculation gives the answer for t channel amplitude (see appendix A for details):
\beq
W^{\mu\nu}\sim \int \frac{d^2k_t}{(2\pi)^2}\frac{J^\mu J^\nu+\kappa_{t\mu}\kappa_{t\nu}}{(k_t^2+x_1x_2Q^2)^2},
\eeq
where 
\begin{eqnarray}
J^\mu &=&2x_1x_2q^{'\mu}-2\vec k_{1t}^2p^\mu+(x_2-x_1)k_{1t}^\mu=J_L^\mu+J_T^\mu \nonumber\\[10pt]
\kappa_\mu&=&\frac{2i}{s}\epsilon_{\mu a b c}k_{1t}^aq^b p^c,J_T^\mu=(x_2-x_1)k_{1t}^\mu,\,\,\, J_L^\mu\sim 2x_1x_2q^{'\mu}-(k^2_t/s)p^\mu .\nonumber\\[10pt]
\label{current}
\end{eqnarray}
Note that $J_T\kappa=0$, i.e. these two transverse currents are actually orthogonal to each other.
\par This is however not the end of the story. It is easy to see that the longitudinal part of the current \ref{current}
is not conserved: $qJ^b_L\ne 0$, where $J^b$ is the contribution to the longitudinal current  of the diagram Fig.3,
i.e. the expression in Eq.\ref{current}.
As it is shown in appendix B,in order to include the gauge invariance one 
has to include the contributions of diagrams 4a and 4c: the full current is a sum $J_L^a+J_L^b+J_L^c$,
where a and c correspond to contributions of the two latter diagrams.

We however do not need the explicit form of the full longitudinal vector current. Indeed it enters the amplitude in the form  $J_Le_L$, where $e_L$ is the  polarisation vector of the longitudinal photon that interacts 
with the hadron. Recall that from  the condition $qe_L=0$ we have $e_L=(q'+xp)/\sqrt{Q^2}=(q+2xp)/\sqrt{Q^2}$.
So the contribution to the amplitude has two parts proportional to $qJ_L$ and $pJ_L$ respectively.  
For the full longitudinal current we have $qJ_L=0$ due to the gauge invariance (the  gauge invariance is explicitly  checked in Appendix B). On the other hand it easy to see that $pJ_L$ is due only to the contribution of the diagram Fig. 3,
for the diagrams Fig. 4a and Fig. 4c  this contribution is zero, since they are proportional to $\gamma^\mu u(p)$ or $\bar u(p)\gamma^\mu$ and these are zero since $\hat pu(p)=0$, due to equations of motion.
Then $e_LJ_L=2xpJ^b_L/Q=2x_1x_2Q$, where $Q=\sqrt{Q^2}$.

\par Then remaining parts of the  calculation is easily done  and give
\beq 
 \frac{d\sigma}{dQ^2dy}=Q_q^2\frac{\alpha_e^2}{2}\int \frac{d^2k_t}{(2\pi)^2}\frac{y}{k_t^2+x_1x_2Q^2)^2}(\vec k_{t}^2(x_1^2+x_2^2)(\frac{(1+(1-y)^2)}{y^2Q^2}
-2m_e^4/Q^4)+8x_1^2x_2^2(1-y)/y^2),
 \eeq
where $W=2kp=ys$ is the squared c.m. energy of e-p collider, and in parton model $x_1+x_2=1$,
and $Q_q$ is the quark charge in electron units,i.e. $Q_u=Q_c=2/3,Q_d=Q_s=-1/3$.
\section{Full DPS cross section and the event number}
\subsection{Hard matrix elements.}
The cross sections  of hard processes $d\sigma/dt$  are usual dijet cross sections calculated with $s\rightarrow  yW$, where $s$ is the invariant energy of the $\gamma p$ system.
.
We have
\beq
d\sigma/dp_{1t}^2=\frac{M^2}{(x_1x_3\sqrt{x_1x_3})16\pi s^{3/2}\sqrt{x_1x_3s-4p_t^2}},
\eeq
where $p_{1t}$ is the dijet transverse scale, and $x_1,x_3$ are jet momentum fractions.
The matrix element $M^2$ of the  quark - gluon scattering is given by \cite{Webber}
\begin{eqnarray}
M^2&=&(4\pi\alpha_s(p_{1t}^2))^2(-\frac{4}{9}(\frac{\hat u}{\hat s}+\frac{\hat s}{\hat u})+\frac{\hat t^2+\hat u^2}{\hat s^2})
\nonumber\\[10pt]
&=&4\pi\alpha_s(p_{1t}^2))^2(-(4/9)(1+z+\frac{1}{1+z})+2(1+z^2))\nonumber\\[10pt]
\label{m1}
\end{eqnarray}
\noindent
where
 \beq \hat s =x_1x_3s, \hat t=-\hat s(1-z)/2, \hat u=-\hat s (1+z)/2, z=\cos\theta=\tanh(y_1-y_3)/2=\sqrt{1-4p_{1t}^2/(x_1x_3s)}.
\eeq
The angle $\theta$ is the scattering angle in the c.m. frame of the dijet.
The region of integration is given by
\beq
x_1x_3s-4(p_t^2+m_c^2)\ge 0,x_1>0.2
\eeq
The integration over the second dijet event goes in the same way, with $x_1\rightarrow x_2,x_3\rightarrow x_4$.

\subsection{Inclusion of a target}
There can be p or A target and the corresponding factor was calculated in \cite{BS},
here we just recall the results.
\subsubsection{The $\gamma p$ case.}
\par In order to estimate whether it is feasible to observe the  MPI events discussed in the previous section, we have to calculate the double differential
cross section and then to convolute it with the photon flux.

 For the case of the proton target
 we have
\beq
\frac{d\sigma}{dx_1x_2dx_3dx_4dp_{1t}^2dp_{2t}^2}=D(x_1,x_2,p_{1t}^2,p_{2t}^2)G(p_{1t}^2,x_3)G(p_{2t}^2,x_4)\frac{d\sigma}{dt_1}\frac{d\sigma}{dt_2}\int \frac{d^2\Delta}{(2\pi)^2} U(\Delta).
\eeq
Here we carried the integration over the momenta $\Delta$ conjugated to the distance between partons,
 obtaining the last multipliers in the equations above. This integral measures the parton wave function at zero
 transverse separation between the partons   and hence it is sensitive to short-range parton-parton correlations.
  For $\gamma p$ case the factor $U(x_1,x_2,\Delta)$, in the approximation when two gluons are not correlated, is equal to a product of two gluon form factors of the proton:
 \beq
  U(\Delta,x_3,x_4)=F_{2g}(\Delta,x_3) F_{2g}(\Delta,x_4).
  \label{U}
  \eeq

 For the numerical estimates we use  the following approximation for $_2GPD$ of the nucleon:
 \beq
 _2D(x_3,x_4,p_{1t}^2,p_{2t}^2,\Delta)=G(x_3,p_{1t})G(x_4,p_{2t})F_{2g}(\Delta,x_3)F_{2g}(\Delta,x_4)
 \eeq
 where the two gluon form factor 
 \beq
 F_{2g}(\Delta)=\frac{1}{(1+\Delta^2/m^2_g)^2}
 \eeq
 and the parameter
 \beq
 m_{g}^2=8/\delta ,
 \eeq
 where
 \beq
 \delta=max(0.28fm^2, 0.31fm^2+0.014fm^2\log(0.1/x)),
 \eeq
 and was determined from the analysis of the exclusive $J/\Psi$ diffractive photoproduction\cite{Frankfurt}.
 The functions $G$ are the gluon pdf of the proton, that we parameterize using
\cite{GRV}.
 Then
 \beq \int \frac{d^2\Delta}{(2\pi)^2}U(\Delta)=\frac{1}{4\pi}\frac{m^2_g(x_3)m^2_g(x_4)(m_g^4(x_3)-m_g^4(x_4)+
 2m^2_g(x_3)m^2_g(x_4)\log(m^2_g(x_4)/m^2_g(x_3))}{(m^2_g(x_3)-m^2_g(x_4))^3}
.
 \eeq
 In the limit $x_3\sim x_4$ we recover
 \beq \int \frac{d^2\Delta}{(2\pi)^2}U(\Delta)=\frac{m^2_g}{(12\pi)}.
 \eeq
 \subsubsection{The $\gamma A$ case.}
 The general expressions for a nuclear target is
\beq
\frac{d\sigma}{dx_1x_2dx_3dx_4dp_{1t}^2dp_{2t}^2}=D(x_1,x_2,p_{1t}^2,p_{2t}^2)G(p_{1t}^2,x_3)G(p_{2t}^2,x_4)\frac{d\sigma}{dt_1}\frac{d\sigma}{dt_2}\int d^2\Delta F'_A(\Delta,-\Delta )
\eeq
 where
\beq
F'_A(\Delta,-\Delta )=F_A(\Delta,-\Delta )+AU(\Delta).
\label{tup}
\eeq
Here $F_A(\Delta,-\Delta )$  is the  nucleus body form factor, and the form factor $U$ was defined in  Eq. \ref{U}. The first term in Eq. \ref{tup}
corresponds to the processes when two gluons
originate
 from the different nucleons in the nucleus
 while the  second term in Eq.\ref{tup} corresponds to the case when they originate
 from the same nucleon. The first term is expected to dominate for heavy nuclei as it scales as
  $A^{4/3}$ \cite{ST,BSW}.
 \par For the nuclear target we have
 \beq
 F_A(\Delta,-\Delta)=F^2(\Delta), F(\Delta)=\int d^2b \exp(i{\vec \Delta}\cdot  {\vec b})T(b),
 \eeq
 where
 \beq
 T(b)=\int dz \rho_A(b,z)dz
 \eeq
 is the nucleus profile function, b is the impact parameter.
 The nuclear form factor integral is expressed through the  profile function as
  \beq G(A)=\int \frac{d^2\Delta}{(2\pi)^2}F(\Delta,-\Delta)=\int T^2(b)d^2b=\pi\int T^2(b)db^2,
  \eeq
  where T(b) is calculated using the conventional mean field nuclear density \cite{AS,Torbard,Vinas}
 \beq
\rho_A(b,z)=\frac{C(A)}{A}\frac{1}{1 + \exp{(\sqrt{b^2 + z^2} - 5.5\cdot A^{1/3})/(2.8)}}.
\eeq
The factor $C(A)$ is a normalization constant
\beq
\int d^2bdz \rho_A(b,z)=A.
\eeq
 Here the  distance scales  are given in  GeV$^{-1}$.
 Note that the nuclear formfactor is independent of $x_3,x_4$.
 \par Then the only difference from the expression for $\gamma p$ collisions for  photon nucleus scattering 
 is substitution 
 \beq
 U(x_3,x_4)\rightarrow AU(x_3,x_4)+G(A)
 \eeq

\subsection{Account of gluon radiation and full expression.}

\par We can now take into account the gluon radiation in the leading logarithmic approximation (LLA) depicted in Fig5.

\begin{figurehere}
\begin{center}
\includegraphics[scale=0.3,angle=90]{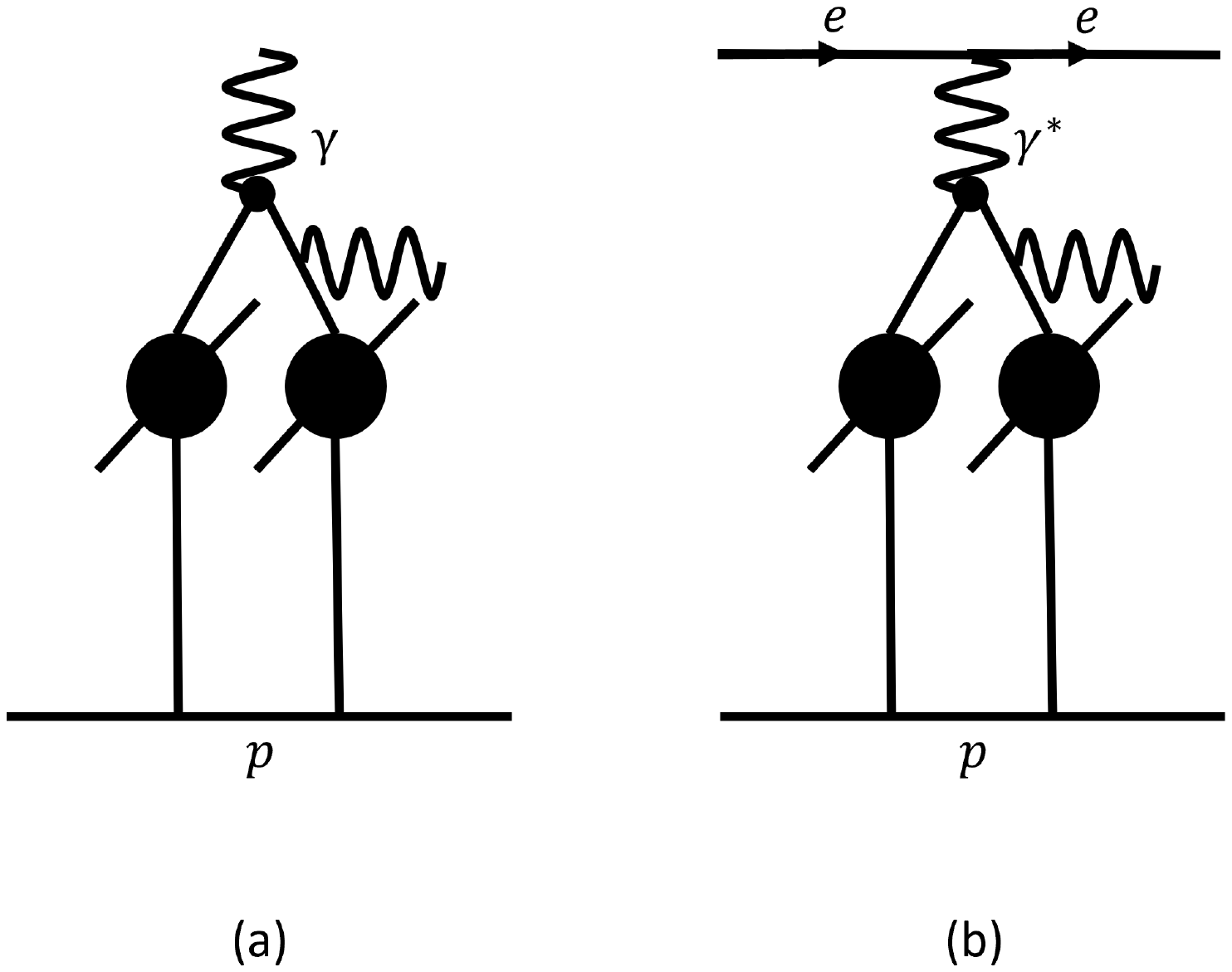}
\end{center}
\label{ran33}
\caption{ DPS with direct photons in pQCD.}
\end{figurehere}
. We  include only the logarithmically 
enhanced DGLAP type diagrams both for the longitudinal and transverse gluons. The DGLAP evolution is easily accounted in $pA=0$ light-cone gauge, see \cite{BDFS3,BDFS4} for details.
We  obtain  the expression for the total number of events for a unit of time
\beq
N=L\int  dk_t^2dzdQ^2dydx_1dx_2dx_3dx_4 dp_{1t}^2dp_{2t}^2 \frac{d\sigma}{dQ^2dydx_1dx_2dx_3dx_4 dp_{1t}^2dp_{2t}^2} 
\eeq
where L is the corresponding luminosity. We have the following conditions determining the integration phase space::
\par in $p_{1t},p_{2t}$-- 
\beq
p_{1t}^2<x_1x_3yW/4, p_{2t}^2<x_2x_4yW/4,
\label{ct1}
\eeq
\beq
 x_1<z<1-x_2
 \label{ct2}
 \eeq
\par
 \beq\frac{Q^2}{W}<y<(Q^2/(2m_e^2)(-1+\sqrt{1+4m_e^2/Q^2})\eeq
coming from $Q^2/(yW)<1$, $Q^2>m_e^2y^2/(1-y)$,\cite{Frixione:1993yw}.
These kinematic constraints  together determine the integration region in the phase space.
For the differential cross-section we then have:
\begin{eqnarray}
 \frac{d\sigma}{dQ^2dydx_1dx_2dx_3dx_4 dp_{1t}^2dp_{2t}^2}
&=&\frac{\alpha_e^2N_c}{2}\int^{ 1-x_2}_{x_1}dz \int \frac{d^2k_t}{(2\pi)^2}\frac{y}{(k_t^2+x_1x_2Q^2)^2}\nonumber\\[10pt]
&\times&(\vec k_{t}^2(z^2+(1-z)^2)(\frac{(1+(1-y)^2)}{y^2Q^2}
-2m_e^4/Q^4)
+8z^2(1-z)^2(1-y)/y^2)\nonumber\\[10pt]
&\times&G_{qA}(x_1/z,k_t^2,p_{1t}^2)G_{\bar qB}(x_2/(1-z),k_t^2,p_{2t}^2)\frac{1}{z(1-z)}\nonumber\\[10pt]
&\times&\frac{M^2}{(x_1x_3\sqrt{x_1x_3})16\pi (yW)^{3/2}\sqrt{x_1x_3yW-4p_{1t}^2}}\nonumber\\[10pt]
&\times&\frac{M^2}{(x_2x_4\sqrt{x_2x_4})16\pi (yW)^{3/2}\sqrt{x_1x_3yW-4p_{2t}^2}}\nonumber\\[10pt]
&\times& U(x_3,x_4)f(x_3,p_{1t}^2)f_g(x_4,p_{2t}^2)\nonumber\\[10pt]
\label{qcd}
\end{eqnarray}
where the hard matrix element $M^2$ is given by Eq. \ref{m1}, functions f are conventinal gluon PDFs \cite{GRV}, and U is a nonperturbative
factor proportional determined in the previous subsection after integration in $\Delta^2$, and proportional to $m^2_g$ for proton target, $\bar q q$ is the quark-antiquark pair created in a photon splitting, and A,B are the resulting jet flavours.

\par The functions $G_{qA},G_{qB}$  are the fundamental solutions of DGLAP equations that correspond to evolution from the scale
$k_t^2$ to the scale $p_{1t}^2$ and $p_{2t}^2$ respectively. If we neglect the evolution (i.e. gluon radiation from the quark model)
these fundamental solutions become delta functions $\delta(1-x_1/z),\delta(1-x_2/(1-z))$, and we obtain from Eq. \ref{qcd} the parton model expression.
\section{Numerics}
\subsection{ Photoproduction revisited.}
\par Let us  first consider photoproduction. We depict the photoproduction due to direct photons in Fig. 6: left- for the case of light quark jet, and right for the case of the charmed quark jet. The two quark jets come from the direct photon side
and two gluon jets come from the nucleon side.
\begin{figurehere}
\begin{center}
\centering
\begin{minipage}{0.49\textwidth}
\centering
    \includegraphics[angle=90,scale=0.35]{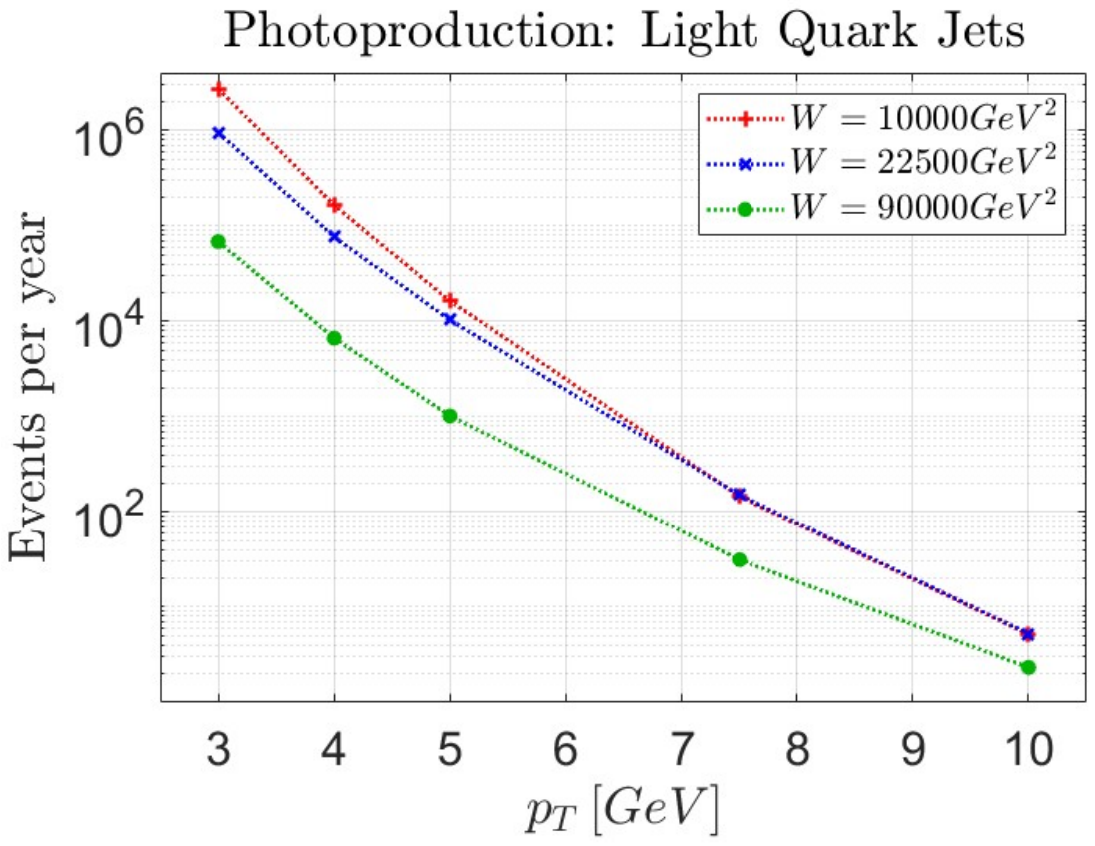}
  \end{minipage}
  \begin{minipage}{0.49\textwidth}
  \centering
    \includegraphics[angle=90,scale=0.35]{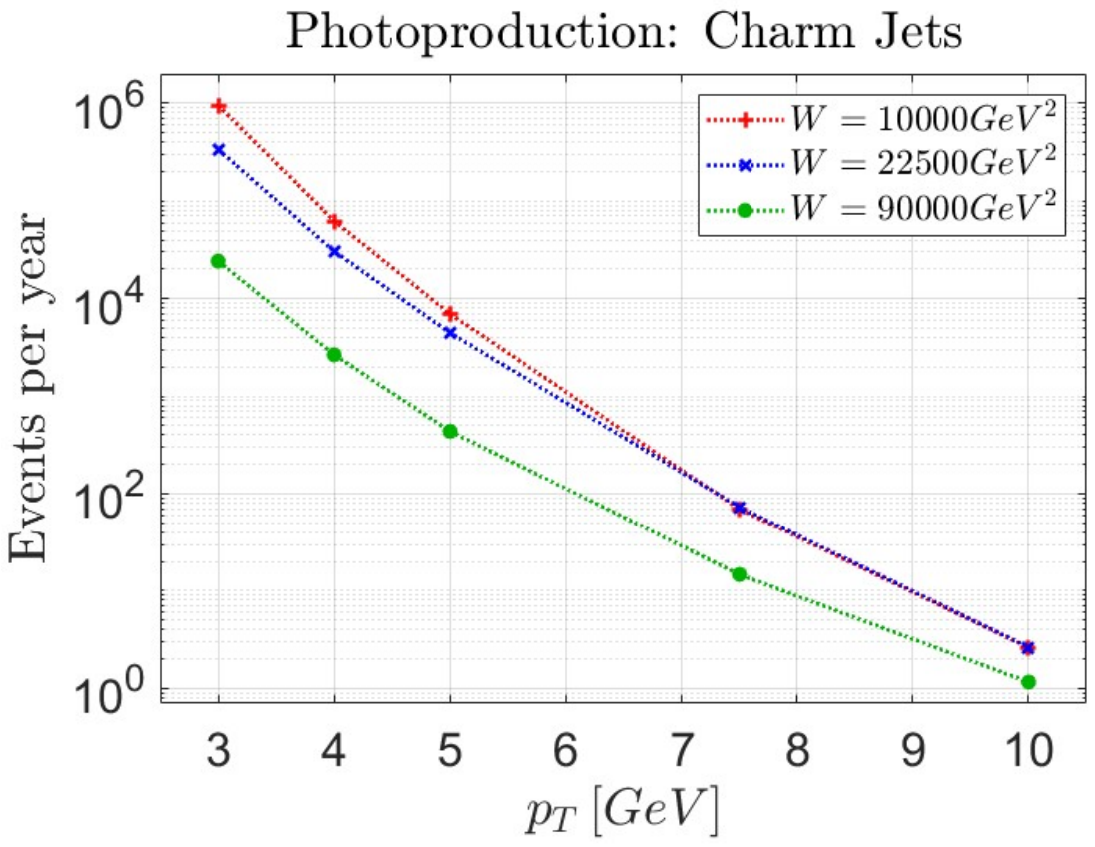}
 \end{minipage}
\end{center}
\label{ran5}
\caption{The  DPS processes with direct photons for  $\gamma p$ deep inelastic  collisions: left-- two light quark and two gluon jets, right-- two charmed jets and two gluon jets as final states}
\end{figurehere}

\

We consider the 3 cases, with different values of W:
 HERA with  $W=90000 $ GeV$^2$, and electron-ion collider. We use two regymes for   IEC:  $W=10000$ GeV$^2$ and maximum,$W=22500$ GeV$^2$.  In the figures we calculate a number of DPS events, that occur for $x_1>0.2,x_2>0.2$ in the logarithmic scale. For the case of light quark jet we sum contributions of u,d and s quarks giving a multuplier 3/2, relative to
  the charm quark jets.
 We assume the luminocities $L=1.7*10^{31} {\rm sm}^{-2}{\rm s}^{-1}$ for HERA,  $L=10^{34} {\rm sm}^{-2}{\rm s}^{-1}$ for IEC and $W=10000$ GeV$^2$, $L=10^{33} {\rm sm}^{-2}{\rm s}^{-1}$ for IEC and $W=22500$ GeV$^2$.
 \par We see that due to higher luminocities the biggest number of DPS events occurs at IEC at $W=10000$ GeV$^2$,
 and this is even larger number than in HERA, i.e. increase in luminocity overcomes the decrease of a number of events due to decrease of the center of mass energy W and the phase space available for hard processes.
 \par For $p_t=3 $ GeV for example we obtain $2*10^6$ events per year for light quark jets, and $\sim 10^6$ for charmed jets
 (i.e. final state with 2 charmed and two gluonic jets), while for HERA we have $3.6*10^5,1.2*10^5$ per year respectively.
 
 \par We also consider in Fig. 7 the change of the number of events if we increase $x_1,x_2$and $x_{\gamma}$ . We see that the decrease is rather slow with an increase of $x_{\gamma}$, similar to the results for such ratio in in \cite{BS}.
 \begin{figurehere}
\begin{center}
\centering
\begin{minipage}{0.49\textwidth}
\centering
    \includegraphics[angle=90,scale=0.35]{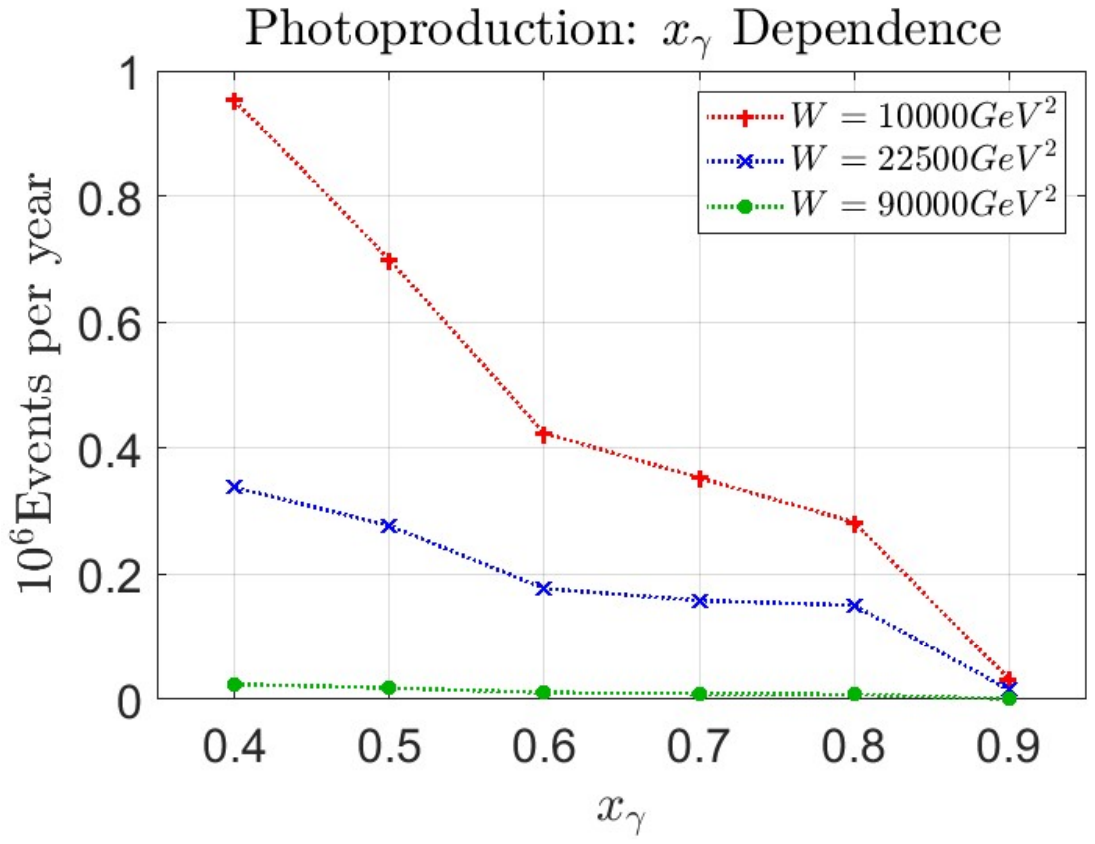}
  \end{minipage}
  \begin{minipage}{0.49\textwidth}
  \centering
    \includegraphics[angle=90,scale=0.35]{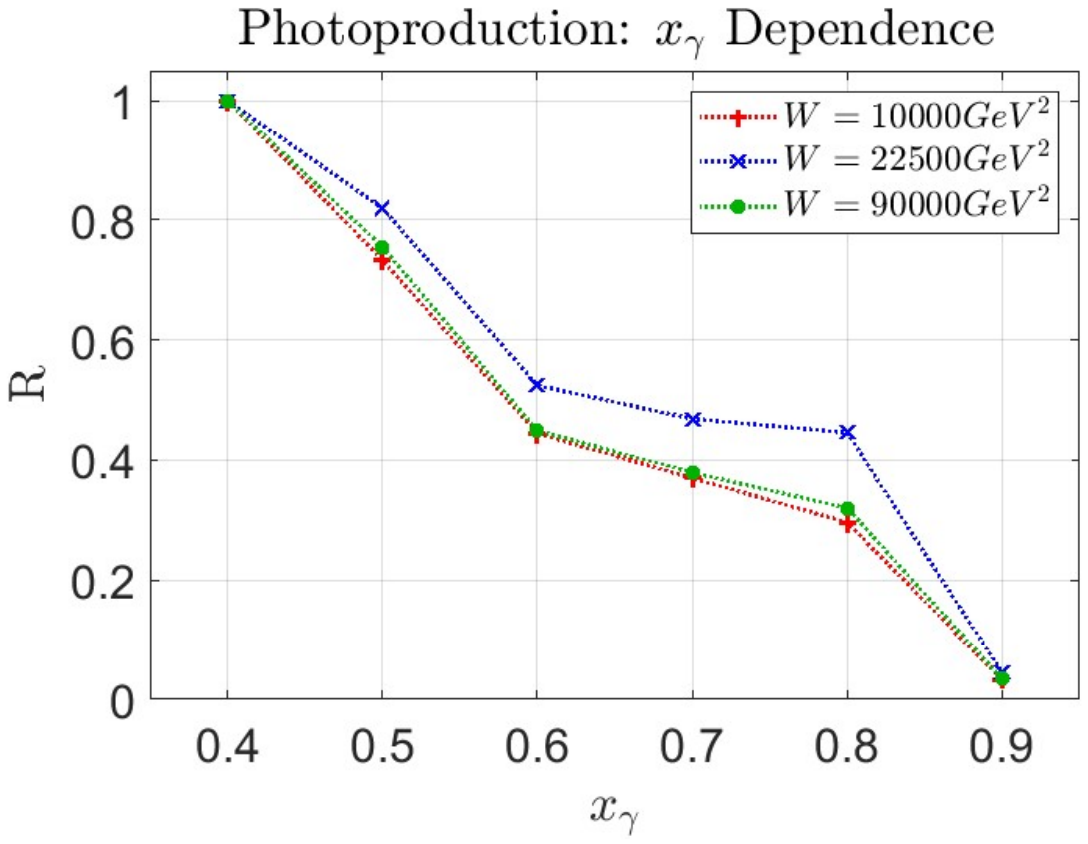}
 \end{minipage}
\end{center}
\label{ran5a}
\caption{The  DPS processes with direct photons for  $\gamma p$ deep inelastic  collisions, charmed jets
dependence of the events number on $x_{\gamma}=x_1+x_2$. $R=\frac{N(x_\gamma)}{N(0.4)}$, 
where N is a number of events for  $x_\gamma$ calculated for $x_1=x_2=x_\gamma/2$.
:Left-- absolute number of events per year, right-- the ratio R defined above.
(For this figure we considered  $x_1=x_2=x_{\gamma}/2$
and the case of two charmed jets
 and two gluon jets as final states, however we expect
that similar dependence on $x_\gamma$ will continue for other final states and for electroproduction.}
\end{figurehere}
 \par In our numerical calculations we used mathematica to integrate over $k_t,z,y$ and then used Simpson rule to integrate 
 over $x_i$ and $p_t$. We integrate over $k_t^2$ starting from $Q_0^2\sim m_\rho^2\sim 4 m_{\rm const}^2\sim 0.6 $ GeV$^2$.
 where $m_{const}$ is the constituent mass of the light quark. The states with small $k_t$ split must be considered 
 as nonperturbative and related to resolved photon contributions.

\par Note that when $x_1+x_2$ become close to 1, the phase space for gluon radiation starts to be extremely small:
$\sim Q_1^2(1-x_1-x_2)$, i.e. the transverse momenta of radiated gluons become small and the coupling constant large,
i.e. the process becomes nonperturbative, and we come close to Landau pole. The situation seems to be close 
what we encounter in the $x\rightarrow 1$ limit in PDF/Drell-Yan processes, where in analogous way the phase space available for radiating gluons becomes $Q^2(1-x)$, and the transverse momenta pf the radiated glions are $k^2_t\le Q^2(1-x)$.
It was shown in \cite{Ciafaloni,amati} that the resummation of such gluons effectively corresponds to the change in the running coupling constant from $\alpha_s(Q^2)$ to $\alpha_s(Q^2(1-x))$, meaning that when $Q^2(1-x)\sim \Lambda_{QCD}^2$, 
we hit the Landau pole, and the perturbative QCD method due to DGLAP become unapplicable.  It is well known that 
the way to get rid of the Landau pole is to carry the so called resummation procedure, which essentially means adding higher twist contributions to get rid of the Landau Pole \cite{catani}. Our case is more complicated since we need to carry
the resummation for 3 point function. Consequently, in order not to hit the Landau pole,
while calculating the number of events  we imposed the kinematic constraint $Q^2(1-x_1-x_2)\ge \Lambda^2_{QCD}$.

\subsection{Electroproduction}
\par For the case of electroproduction we consider both the charm and light quark jets. We take the
transverse scales of the jets for the examples we made the numerical estimates to be 3 and 5 GeV.
\begin{figurehere}
\begin{center}
\centering
\begin{minipage}{0.49\textwidth}
\centering
    \includegraphics[angle=90,scale=0.35]{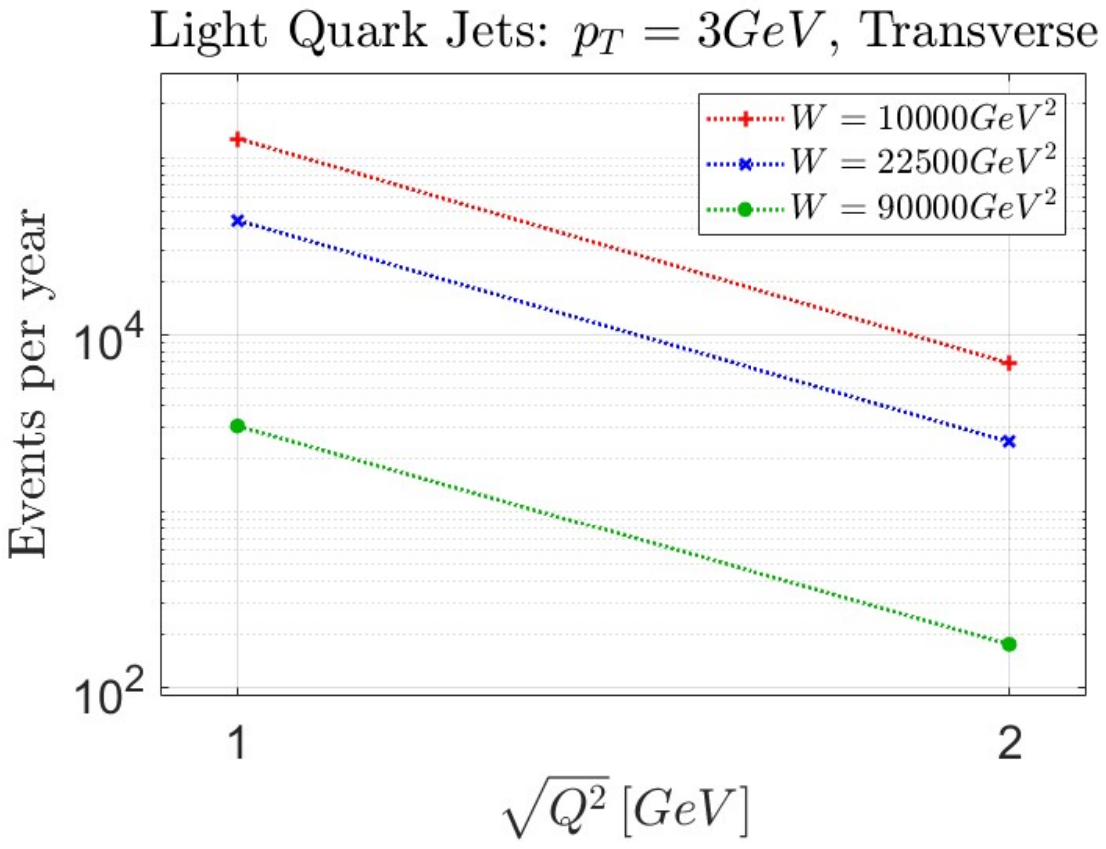}
  \end{minipage}
  \begin{minipage}{0.49\textwidth}
  \centering
    \includegraphics[angle=90,scale=0.35]{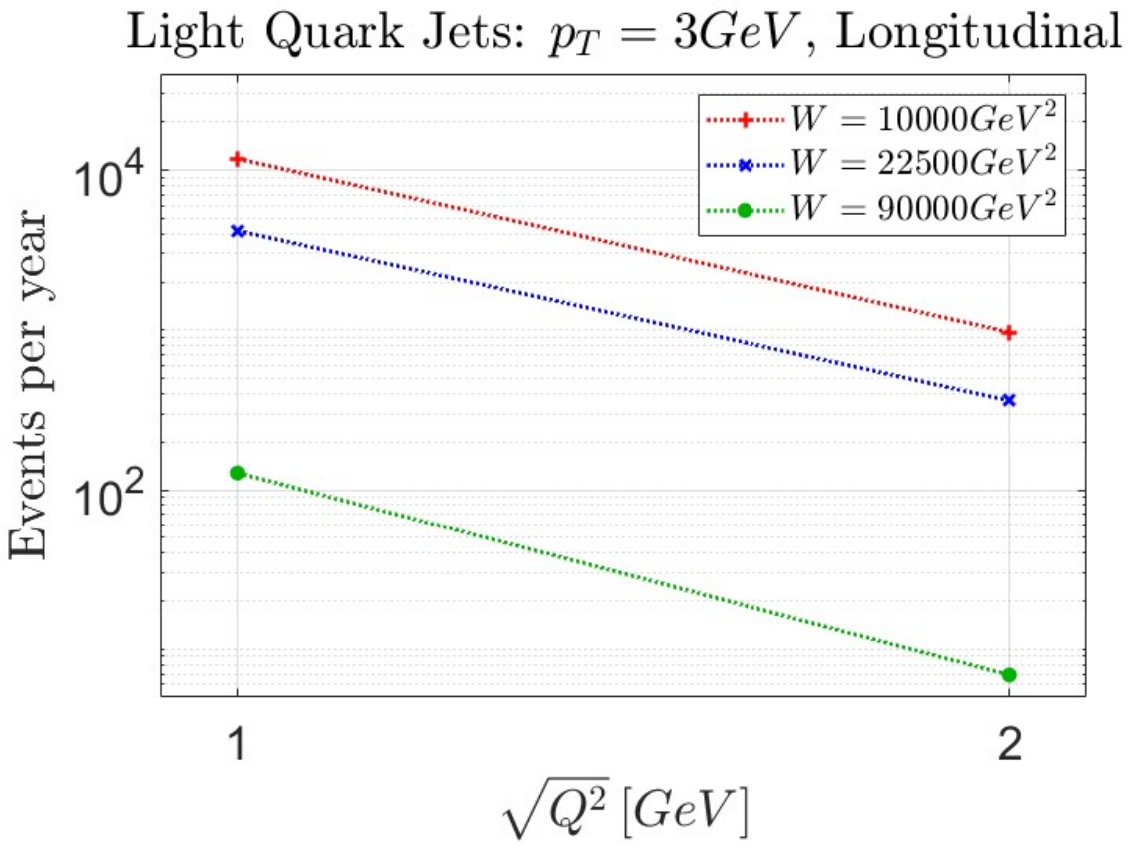}
 \end{minipage}
\end{center}
\label{ran6}
\caption{The  DPS proesses with direct photons for  $\gamma p$ deep inelastic  collisions as function of Q for fixed $p_T$: 
light quark jets with $p_T=3$ GeV,electroproduction, left-transverse photon contribution, right-longitudinal photons}
\end{figurehere}

\begin{figurehere}
\begin{center}
\centering
\begin{minipage}{0.49\textwidth}
\centering
    \includegraphics[angle=90,scale=0.35]{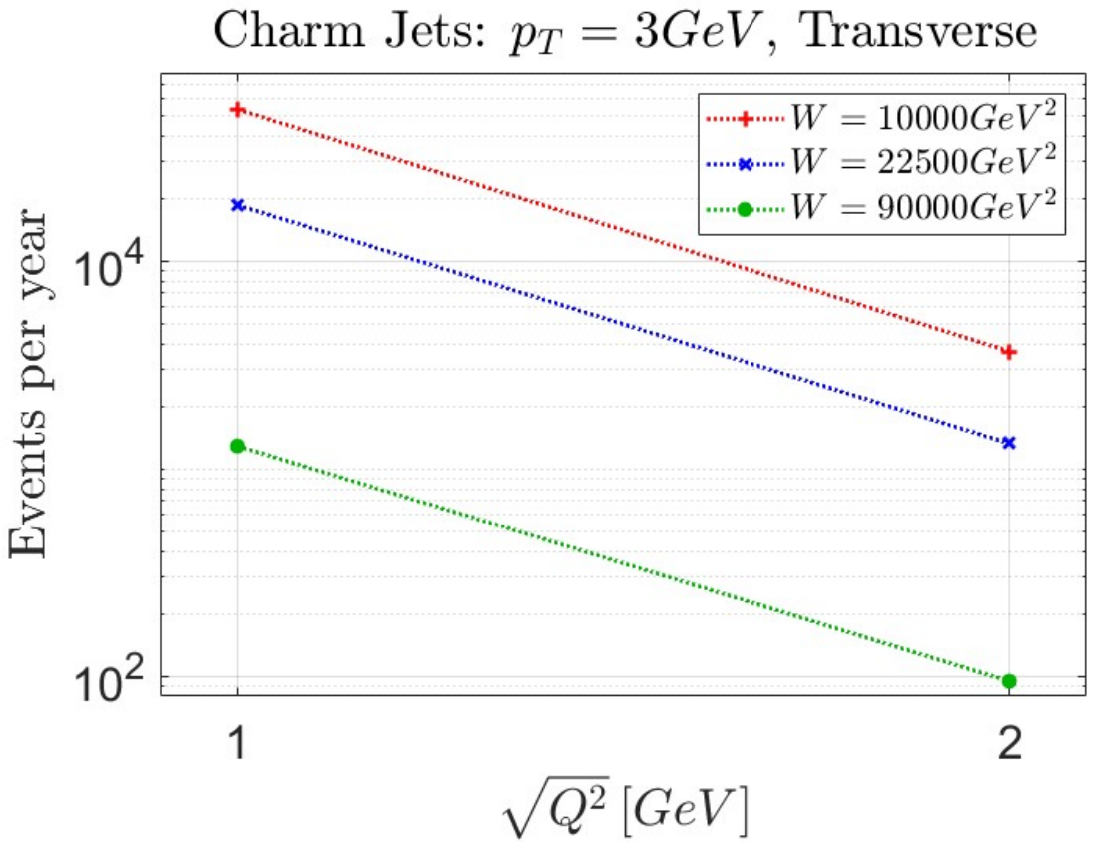}
  \end{minipage}
  \begin{minipage}{0.49\textwidth}
  \centering
    \includegraphics[angle=90,scale=0.35]{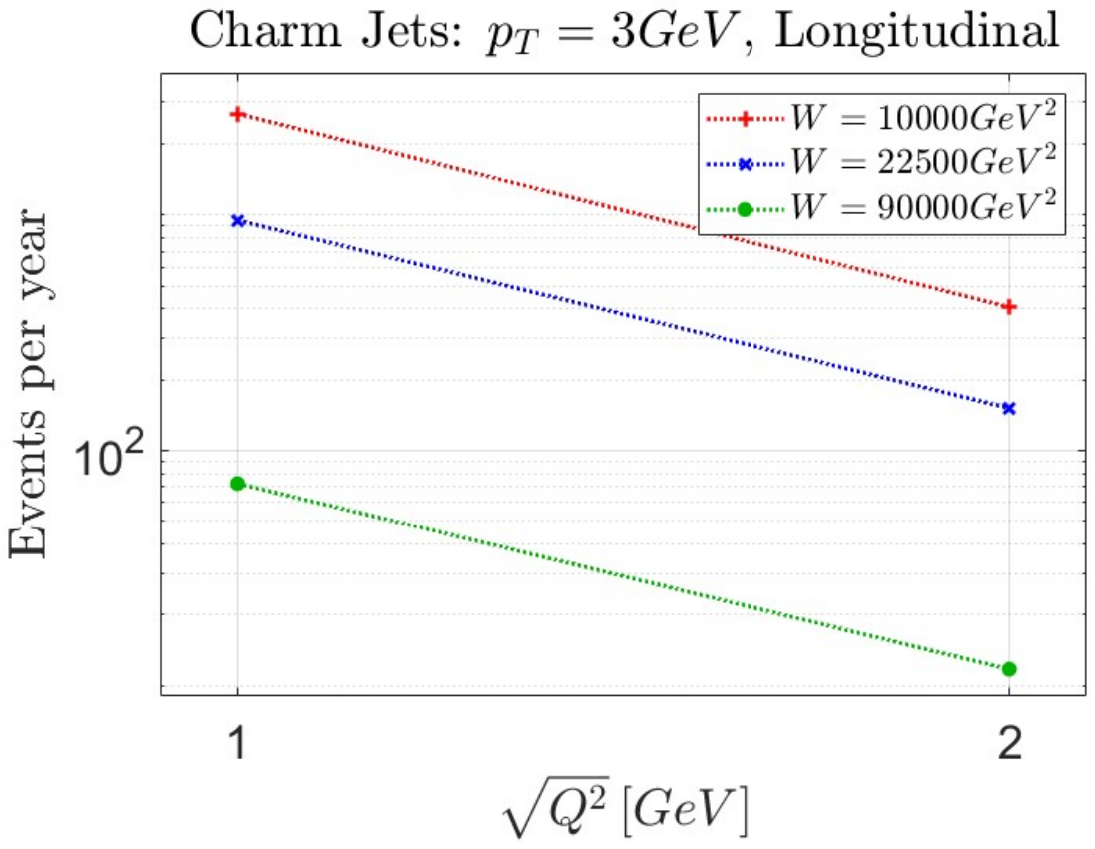}
 \end{minipage}
\end{center}
\label{ran7}
\caption{The  DPS proesses with direct photons for  $\gamma p$ deep inelastic  collisions as function of Q for fixed $p_T$: 
charm quark jets with $p_T=3$ GeV,electroproduction, left-transverse photon contribution, right-longitudinal photons}
\end{figurehere}

\begin{figurehere}
\begin{center}
\centering
\begin{minipage}{0.49\textwidth}
\centering
    \includegraphics[angle=90,scale=0.35]{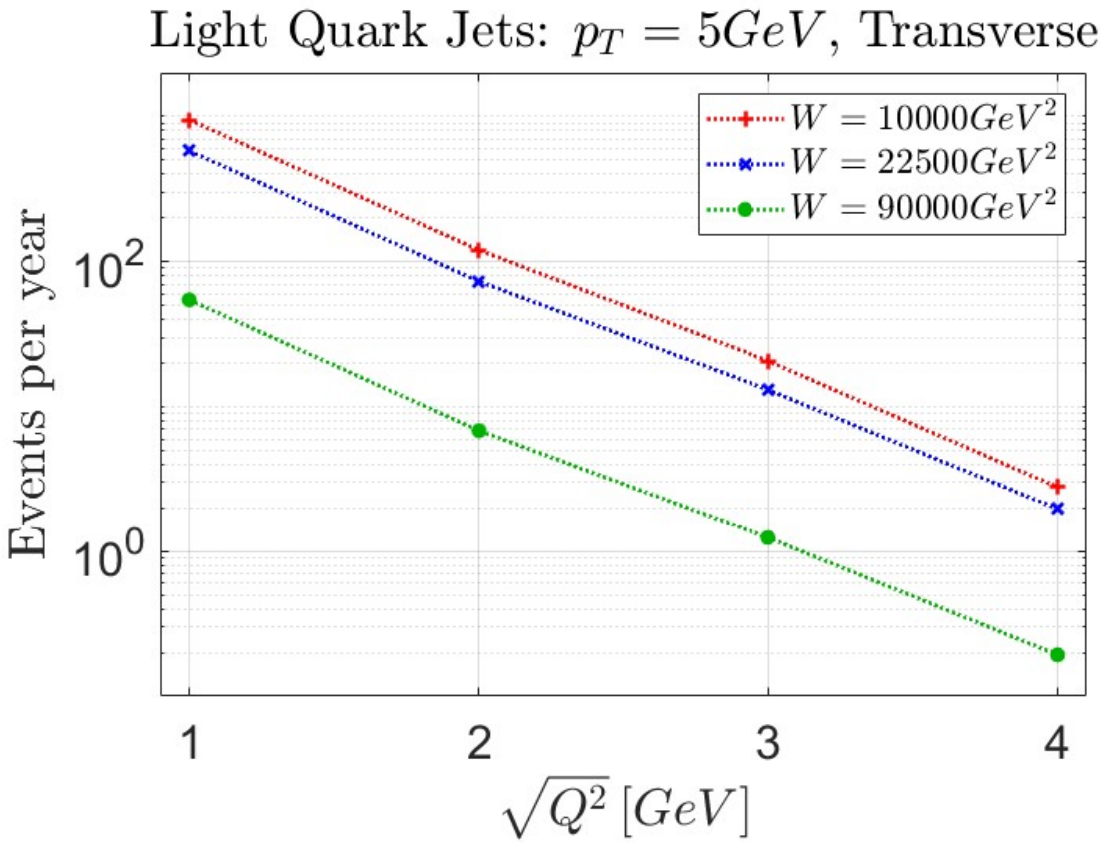}
  \end{minipage}
  \begin{minipage}{0.49\textwidth}
  \centering
    \includegraphics[angle=90,scale=0.35]{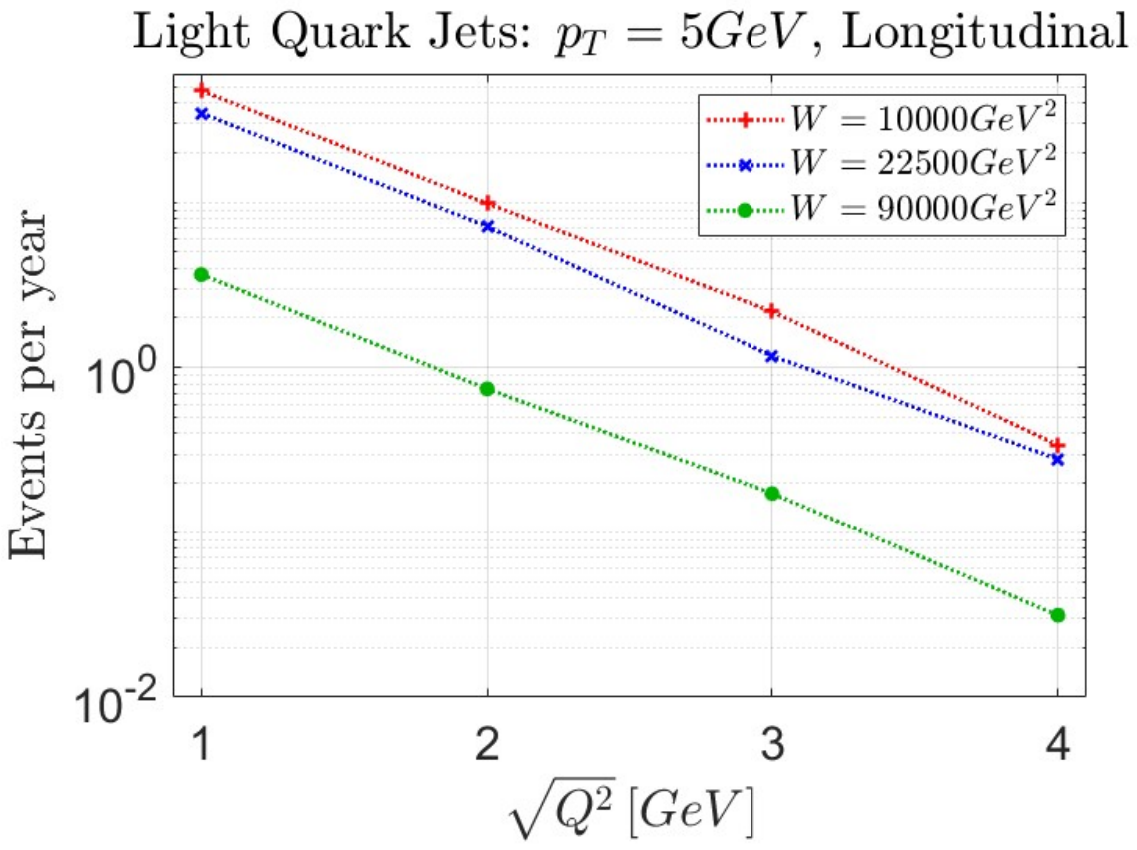}
 \end{minipage}
\end{center}
\label{ran8}
\caption{The  DPS proesses with direct photons for  $\gamma p$ deep inelastic  collisions as function of Q for fixed $p_T$: 
light quark jets with $p_T=5$ GeV,electroproduction, left-transverse photon contribution, right-longitudinal photons}
\end{figurehere}

\begin{figurehere}
\begin{center}
\centering
\begin{minipage}{0.49\textwidth}
\centering
    \includegraphics[angle=90,scale=0.35]{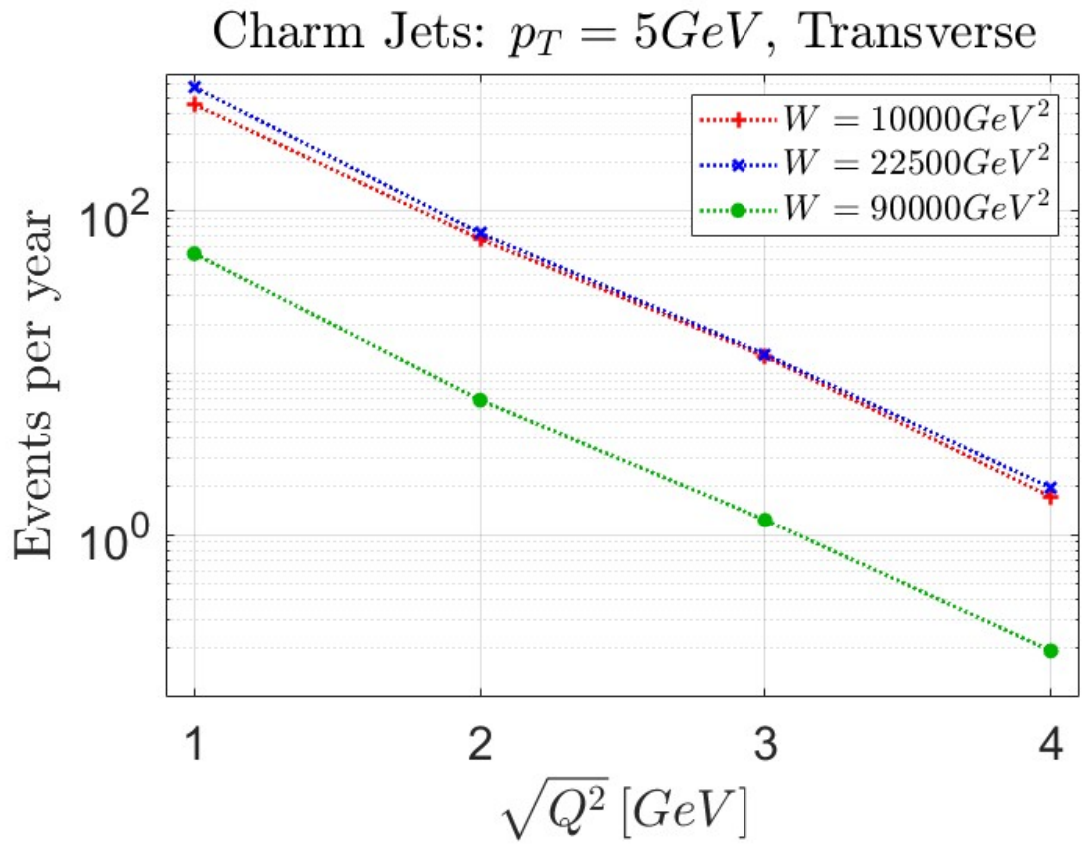}
  \end{minipage}
  \begin{minipage}{0.49\textwidth}
  \centering
    \includegraphics[angle=90,scale=0.35]{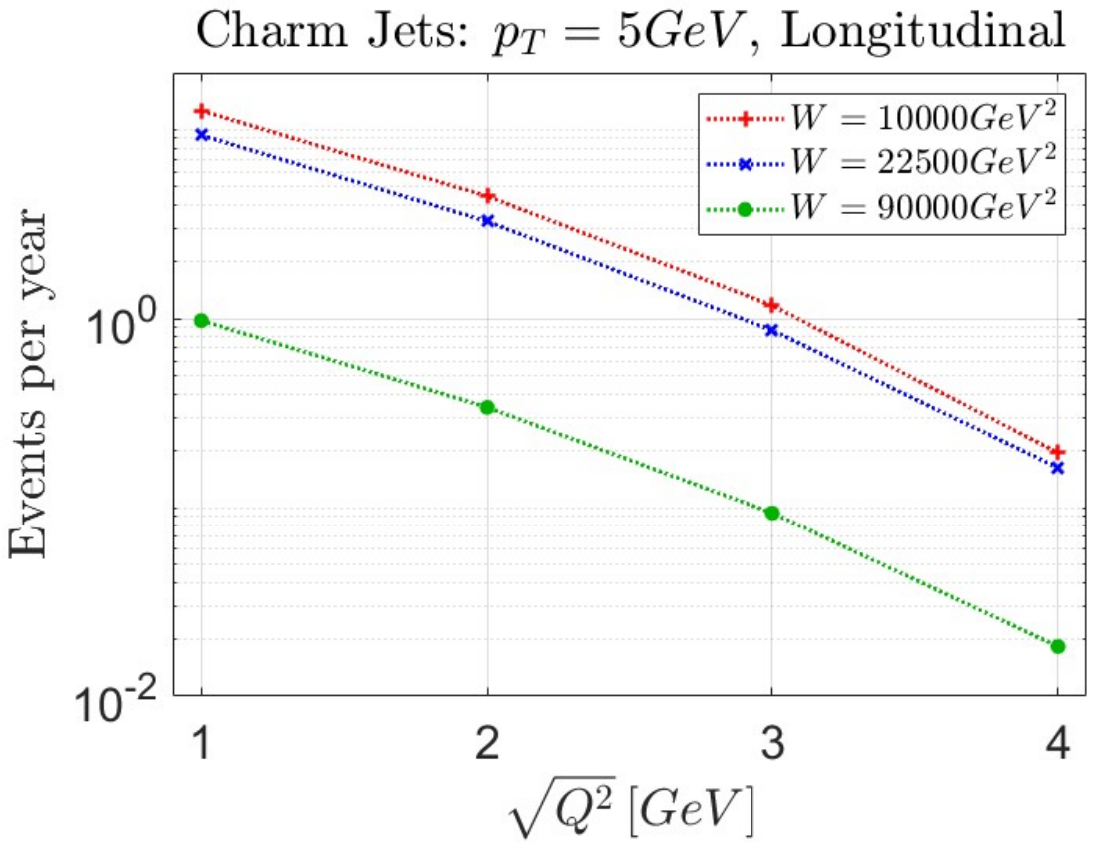}
 \end{minipage}
\end{center}
\label{ran9}
\caption{The  DPS proesses with direct photons for  $\gamma p$ deep inelastic  collisions as function of Q for fixed $p_T$: 
charm quark jets with $p_T=5$ GeV,electroproduction, left-transverse photon contribution, right-longitudinal photons}
\end{figurehere}

 In Figs.8-11  we  depict the DPS event numbers for HERA and IEC,
with the same parameters as in the previous subsection, as functions of the photon virtuality Q. We consider separately the contributions of the longitudinal and transverse photon. We see however that in all considered examples the contribution of the longitudinal photons is rather small only 5-10 $\%$ at most relative to transverse photon contribution, consequently in these 
examples total number of DPS is approximately equal to the transverse photon contribution.
Note that in this case we integrate over $k^2_t\ge Q^2$, where Q is the momentum of the virtual photon, and effectively
the invariant mass of the initial quark-antiquark pair. We also see that the number of DPS in electroproduction
is significantly smaller than the number of DPS events in photoproduction.
\subsection{Photon-Nucleus reactions.}
The factor $G(A)$ for A=208 is equal to G(A)=11.5, on the other hand for A=208 $AU(x_3,x_4)$ varies from 4.5 to 5.7
for different $x_3,x_4$ relevant for the integration region for HERA, while for IEC the relevant region is even smaller
from  5.3 to 5.7 for W=10000 GeV$^2$, and  5.1 to 5.7,  thus the corresponding multiplier for A=208 (Pb)
is for IEC 16.9-17.3 and 16.7-17.3, in other words with good accuracy  a number of events in photon-nucleus collisions at IEC and HERA is 3 times larger than for photon-proton DPS events, for the whole region of $x_1,x_2$ considered here
( and taken for the same luminocity).
\section{Conclusion.}\par We found that a rather large number of DPS events
of order $10^4-10^5$ is due to direct photons in the IEC  in different kinematics and in HERA,  for both  $\gamma p$ and $\gamma A$ reactions. We discussed how the gauge invariance is realized and can be used to calculate cross section due to longitudinal gluons.
The study of these DPS like  events will permit  us to study the specific properties of \12 mechanism, and of the 
short range transverse correlations in the nucleon. The DPS events in the limit $x_1+x_2\rightarrow 1$ deserve further study
since they may be controlled by very soft gluons, and one needs to develop the resummation procedure
along the lines of \cite{catani}  to take them into account. 
\ \acknowledgements We thank M. Strikman and Yu. Dokshitzer for very usefull discussions,
 and M. Strikman for reading this article. This work was supported by BSF grant  2033344.
\appendix
\section{Breit system calculation for DPS.}
\par In this appendix we show the details of the calculation of the  electroproduction cross section for DPS. We shall carry
the calculation in Breit system
\cite{Jezuita-Dabrowska:2002tjg}.
\par \beq
q=Q(0,0,0,1),p=E_p(1,0,0,-1),
\eeq
where $E_p$ is the energy of the proton, $s=2E_pQ$.
\par In this system we can parametrize  the momenta of initial and scattered electrons as 
\begin{eqnarray}
k&=&\frac{1}{2}Q(\cosh{(\psi)},\sinh{(\psi)}\cos{(\phi_e)},\sinh{(\psi)}\sin{(\phi_e)},1)\nonumber\\[10pt]
k'&=&\frac{1}{2}Q(\cosh{(\psi)},\sinh{(\psi)}\cos{(\phi_e)},\sinh{(\psi)}\sin{(\phi_e)},-1)\nonumber\\[10pt]
\label{k1}
\end{eqnarray}
If $y=qp_p/kp_p$, then
\beq
\cosh(\psi)=(2-y)/y,\,\,\,\, \sinh (\psi )=2\sqrt{1-y}/y
\label{santa}
\eeq

The leptonic tensor is
\beq
L_{\alpha\beta}\sim 4(k^\mu k^{'\nu}+k^\nu k^{'\mu}-g^{\mu\nu}Q^2/2)
\eeq
The emitted photon polarization vectors are
\begin{eqnarray}
e^1_T&=&(0,\cos(\phi_\gamma),\sin (\phi_\gamma),0)\nonumber\\[10pt]
e^2_T&=&(0,-\sin (\phi_\gamma),\cos(\phi_\gamma),0)\nonumber\\[10pt]
e_L&=&(1,0,0,0)\nonumber\\[10pt]
\end{eqnarray}
where we chooze linearly polarized transverse photons as a basis.
It is easy to calculate 
\begin{eqnarray}
L^{\alpha\beta}e_{L\alpha} e_{L\beta}&=&2Q^2(1-y)/y^2\nonumber\\[10pt]
L^{\alpha\beta}e_{T1\alpha} e_{T1\beta}&=&\frac{Q^2}{2}(1+2(ke^1_t)^2)=\frac{Q^2}{2}(1+\sinh(\psi)^2\cos(\phi_e-\phi_\gamma)^2)\nonumber\\[10pt]
L^{\alpha\beta}e_{T2\alpha} e_{T2\beta}&=&\frac{Q^2}{2}(1+\sinh (\psi)^2\sin (\phi_e-\phi_\gamma)^2)\nonumber\\[10pt]
L^{\alpha\beta}e_{T1\alpha} e_{T2\beta}&=&\frac{Q^2}{2}\sinh(\psi)^2\sin(2(\phi_e-\phi_\gamma))/2\nonumber\\[10pt]
L^{\alpha\beta}e_{L\alpha} e_{T1\beta}&=&-\frac{Q^2}{2y}\sinh (\psi)\cos(\phi_e-\phi_\gamma)\nonumber\\[10pt]
L^{\alpha\beta}e_{L\alpha} e_{T2\beta}&=&-\frac{Q^2}{2y}\sinh (\psi)\sin(\phi_e-\phi_\gamma)\nonumber\\[10pt]
\end{eqnarray}
Next  we consider the corresponding contractions with the hadronic tensor.
\begin{eqnarray}
Je_L&=&J_Le_L=2x_1x_2Q\nonumber\\[10pt]
Je_T^1&=&J_Te_T^1+i\kappa  e_T^1=-((x_1-x_2)\cos (\phi_q-\phi_\gamma)-i\sin (\phi_q-\phi_\gamma))\nonumber\\[10pt]
Je_T^2&=&J_Te_T^2+i\kappa  e_T^2=--((x_1-x_2)\sin(\phi_q-\phi_\gamma)+i\cos(\phi_q-\phi_\gamma))\nonumber\\[10pt]
\end{eqnarray}
where in Breit system 
\beq
\kappa_{t}^i=\epsilon^{ij}k_{1tj}, \vec k_{1t}=(\cos(\phi_q),\sin(\phi_q)),\vec \kappa =(-\sin(\phi_q),\cos(\phi_q)
\eeq
Then it is easy to see that L-T cross terms are zero. For part of the  cross section due to longitudinal photons we have
\beq
\sigma_{LL}\sim (L_{\alpha\beta} e_L^\alpha e_L^\beta )(Je_L)^2=8Q^4x_1^2x_2^2(1-y)/y^2
\eeq
For transverse part we have
\beq
\sigma_{TT}\sim  \sum_{\lambda =1,2} L_{\alpha\beta}e_T^{\alpha\lambda}e_T^{\beta\lambda}(Je_T^\lambda)(Je_T\lambda)+2L_{\alpha\beta}e_T^{\alpha 1}e_T^{\beta 2}(Je_T^1)(Je_T^2)
\eeq
This can be rewritten as 
\begin{eqnarray}
\sigma_{TT}&\sim& Q^2(\sinh(\psi)^2\cos(\phi_e-\phi_\gamma)^2+1)(x_1^2+x_2^2)\cos(\phi_q-\phi_\gamma)^2\nonumber\\[10pt]
&+&Q^2(\sinh(\psi)^2\sin(\phi_e-\phi_\gamma)^2+1)(x_1^2+x_2^2)\sin(\phi_q-\phi_\gamma)^2\nonumber\\[10pt]
&-&Q^2(x_1x_2)(\sinh(\psi)^2\sin(2(\phi_q-\phi_\gamma))\sin(2(\phi_e-\phi_\gamma))\nonumber\\[10pt]
\end{eqnarray}
we now average over the photon polarisation angle $\phi_\gamma$. 
\begin{eqnarray}
\int  d\phi_\gamma\cos(\phi_q-\phi_\gamma)^2 &=&(1/2)*2\pi\nonumber\\[10pt]
\int d\phi_\gamma\cos(\phi_q-\phi_\gamma)^2 \cos(\phi_e-\phi_\gamma)^2&=&2\pi*(1/4)(1+\cos(2(\phi_q-\phi_e))/2)\nonumber
\\[10pt]
\int d\phi_\gamma\sin(\phi_q-\phi_\gamma)^2 \sin(\phi_e-\phi_\gamma)^2&=&2\pi*(1/4)(1+\cos(2(\phi_q-\phi_e))/2)\nonumber\\[10pt]
\int  d\phi_\gamma\sin(2(\phi_q-\phi_\gamma))\sin(2(\phi_e-\phi_\gamma))&=&-\pi \cos (2(\phi_q-\phi_e))\nonumber\\[10pt]
\end{eqnarray}
This means that 
\beq
\sigma_{TT}\sim 2Q^2(x_1^2+x_2^2)(1+\sinh(\psi)^2/2)*(1+\cos(2(\phi_q-\phi_e))/2)+Q^2x_1x_2\sinh(\psi)^2\cos (2(\phi_q-\phi_e))/2)
\eeq
We see that the interaction with longitudinal gluons is isotropic  while the transverse gluons have angular correlations:
Averaging over angular correlations we get 
\beq
\sigma_{TT}\sim Q^2\vec k_{1t}^2(x_1^2+x_2^2)(1+\sinh(\psi)^2/2)=Q^2\vec k_{1t}^2(x_1^2+x_2^2)(1+(1-y)^2)/y^2
\eeq
where $\psi$ is determined by Eq. \ref{santa}.
 Thus , including the term with the mass of the electron, we get
 \beq
 \frac{d\sigma}{dQ^2dy}=\frac{\alpha_e^2N_c}{2}\int \frac{d^2k_t}{(2\pi)^2}\frac{y}{(k_t^2+x_1x_2Q^2)^2}(\vec k_{t}^2(x_1^2+x_2^2)(\frac{(1+(1-y)^2)}{y^2Q^2}
-2m_e^2/Q^4)+8x_1^2x_2^2(1-y)/y^2)
 \eeq
 where we have the same normalisation as in \cite{BS} for the total cross section and $\alpha_e=1/137$ is an electromagnetic constant. 
 There are two  kinematic boundaries:
First, since 
 \beq
 x=Q^2/s=Q^2/(yW)<1
 \eeq
 then
 \beq
y>Q^2/W.
\eeq
The second boundary comes from
 \beq
 Q^2>\frac{m^2_ey^2}{1-y}.
 \eeq
 Solving this enequality we obtain:
 \beq
  y\le \frac{Q^2}{2m_e^2}(-1+\sqrt{1+4m_e^2/Q^2})\sim 1, {\rm if}Q^2>>m_e^2.
 \eeq
 So for fixed $Q^2$ we have the limits
 \beq
 \frac{Q^2}{2m_e^2}(-1+\sqrt{1+4m_e^2/Q^2})> y>Q^2/W.
 \eeq
 The latter equations means that 
 \beq
 y\le 1 \,\,Q^2\gg m_e^2;\,\,y\le Q/m_e, Q^2\ll m_e^2
 \eeq
 In addition there is a constraint due to hard processes (see section 3C).
 \section{Gauge invariance}
 \par It is interesting to note that the longitudinal part of the current calculated in the previous section is not conserved .
In order to achieve gauge invariance we need to consider not only the diagram 10b, but all three diagrams in 
   Fig. 10. It is easy to see that 
 diagrams 10a,10c do not contribute to the scattering amplitudes,but  just account for  the gauge invariance.
We want to look at a proton with momentum $p$ interacting with a photon with momentum $q$ such that $q^2=-Q^2<0$ via two quarks with momentum $k_3$ and $k_4$ such that $k_3||k_4||p$, resulting in two dijets with momentums $Q_1$ and $Q_2$. In Fig. \ref{3amps} we present the three amplitudes that contribute to the process and that need to be accounted for to satisfy the Ward-Takahashi identity. 

\begin{figurehere}
\begin{center}
\includegraphics[scale=0.4]{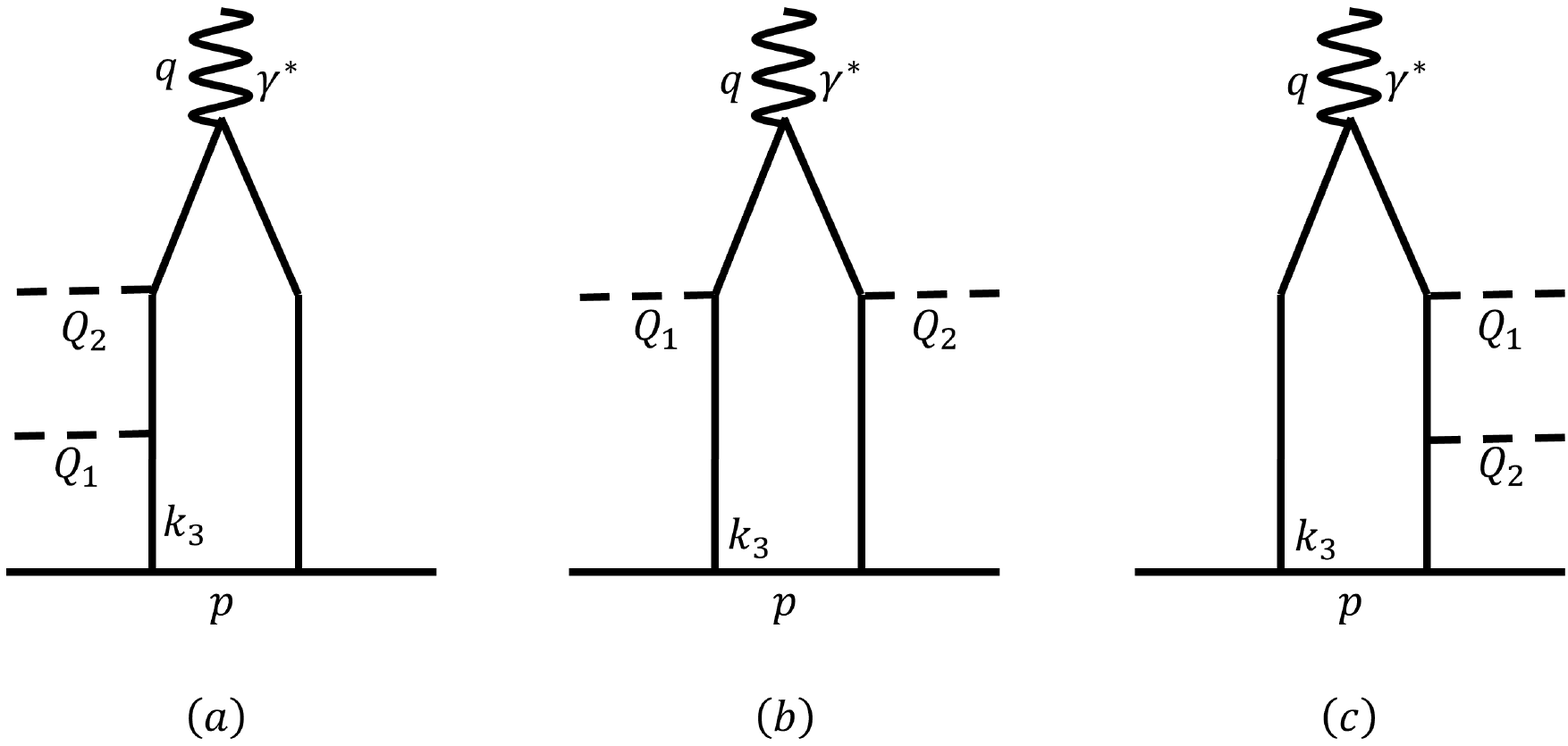}
\end{center}
\label{3amps}
\caption{The three amplitudes that contribute to the process.}
\end{figurehere}

We introduce the center of energy scale $s\equiv2qp$ and the vector $q'\equiv q+xp$ where $x=Q^2/s$, we are going to assume $p^2=0$, and from that we can see that $\left(q'\right)^2=0$. We can parameterize the outgoing momentums as: $Q_1^{\mu}=x_1\left(q'\right)^{\mu}+x_3p^{\mu}+k_t^{\mu},\;Q_1^{\mu}=x_2\left(q'\right)^{\mu}+x_4p^{\mu}-k_t^{\mu}$, where $x_1+x_2=1$ and $k_t$ is perpendicular to both $q'$ and $p$, and $k_3=\left(x_3-\beta\right)p,\;k_4=\left(x_4+\beta+x\right)p$. With this, we can write out the amplitudes as:
\beeq
(a)&\propto&\bar{u}\left(p\right)\frac{-x_{1}\hat{q'}-\beta\hat{p}-\hat{k}_{t}}{x_{1}\beta s-\vec{k}_{t}^{2}+i\epsilon}\frac{-\hat{q'}-\left(x_{4}+\beta\right)\hat{p}}{\left(x_{4}+\beta\right)s+i\epsilon}\gamma^{\mu}u\left(p\right).\nonumber\\[10pt]
(b)&\propto&\bar{u}\left(p\right)\frac{-x_{1}\hat{q'}-\beta\hat{p}-\hat{k}_{t}}{x_{1}\beta s-\vec{k}_{t}^{2}+i\epsilon}\gamma^{\mu}\frac{x_{2}\hat{q'}-\left(\beta+x\right)\hat{p}-\hat{k}_{t}}{-x_{2}\left(\beta+x\right)s-\vec{k}_{t}^{2}+i\epsilon}u\left(p\right)\nonumber\\[10pt]
(c)&\propto&\bar{u}\left(p\right)\gamma^{\mu}\frac{\hat{q'}+\left(x_{3}-\beta-x\right)\hat{p}}{\left(x_{3}-\beta-x\right)s+i\epsilon}\frac{x_{2}\hat{q'}-\left(\beta+x\right)\hat{p}-\hat{k}_{t}}{-x_{2}\left(\beta+x\right)s-\vec{k}_{t}^{2}+i\epsilon}u\left(p\right),
\eeeq
taking all three amplitudes together we can now see what we get if we take a product of them with different vectors, a useful basis we can work with is the set $q',\;p,\;k_t$ and $\kappa$, where $\kappa$ is a vector perpendicular to all the other three and satisfies: $\kappa^2=k_t^2=-\vec{k_t}^2$. We now define the sum as a new "vector":
\beeq
V^{\mu}\left(\beta\right)&=&\bar{u}\left(p\right)\left[\frac{-x_{1}\hat{q'}-\hat{k}_{t}}{x_{1}\beta s-\vec{k}_{t}^{2}+i\epsilon}\frac{-\hat{q'}-\left(x_{4}+\beta\right)\hat{p}}{\left(x_{4}+\beta\right)s+i\epsilon}\gamma^{\mu}\right.\nonumber\\[10pt]
&+&\frac{-x_{1}\hat{q'}-\hat{k}_{t}}{x_{1}\beta s-\vec{k}_{t}^{2}+i\epsilon}\gamma^{\mu}\frac{x_{2}\hat{q'}-\hat{k}_{t}}{-x_{2}\left(\beta+x\right)s-\vec{k}_{t}^{2}+i\epsilon}\nonumber\\[10pt]
&+&\left.\gamma^{\mu}\frac{\hat{q'}+\left(x_{3}-\beta-x\right)\hat{p}}{\left(x_{3}-\beta-x\right)s+i\epsilon}\frac{x_{2}\hat{q'}-\hat{k}_{t}}{-x_{2}\left(\beta+x\right)s-\vec{k}_{t}^{2}+i\epsilon}\right]u\left(p\right)\nonumber\\[10pt],
\eeeq
where the terms of the form $\bar{u}\left(p\right)\hat{p}$ and $\hat{p}u\left(p\right)$ are taken to be zero. And look at the products with the basis we chose:
\beeq
q'V\left(\beta\right)&=&\frac{-x_{1}x_{2}xs^{2}}{\left(x_{1}\beta s-\vec{k}_{t}^{2}+i\epsilon\right)\left(-x_{2}\left(\beta+x\right)s-\vec{k}_{t}^{2}+i\epsilon\right)},\nonumber\\[10pt]
pV\left(\beta\right)&=&\frac{-x_{1}x_{2}s^{2}}{\left(x_{1}\beta s-\vec{k}_{t}^{2}+i\epsilon\right)\left(-x_{2}\left(\beta+x\right)s-\vec{k}_{t}^{2}+i\epsilon\right)},\nonumber\\[10pt]
k_tV\left(\beta\right)&=&\left[\begin{array}{c}
\frac{1}{\left(x_{1}\beta s-\vec{k}_{t}^{2}+i\epsilon\right)\left(\left(x_{4}+\beta\right)s+i\epsilon\right)}\\
+\frac{\left(x_{2}-x_{1}\right)}{\left(x_{1}\beta s-\vec{k}_{t}^{2}+i\epsilon\right)\left(-x_{2}\left(\beta+x\right)s-\vec{k}_{t}^{2}+i\epsilon\right)}\\
-\frac{1}{\left(\left(x_{3}-\beta-x\right)s+i\epsilon\right)\left(-x_{2}\left(\beta+x\right)s-\vec{k}_{t}^{2}+i\epsilon\right)}
\end{array}\right]\vec{k}_{t}^{2}s,\nonumber\\[10pt]
\kappa V\left(\beta\right)&=&\left[\begin{array}{c}
\frac{1}{\left(x_{1}\beta s-\vec{k}_{t}^{2}+i\epsilon\right)\left(\left(x_{4}+\beta\right)s+i\epsilon\right)}\\
+\frac{1}{\left(x_{1}\beta s-\vec{k}_{t}^{2}+i\epsilon\right)\left(-x_{2}\left(\beta+x\right)s-\vec{k}_{t}^{2}+i\epsilon\right)}\\
+\frac{1}{\left(\left(x_{3}-\beta-x\right)s+i\epsilon\right)\left(-x_{2}\left(\beta+x\right)s-\vec{k}_{t}^{2}+i\epsilon\right)}
\end{array}\right]2i\lambda\varepsilon^{abcd}p_{\alpha}\left(q'\right)_{b}\kappa_{c}\left(k_{t}\right)_{d},
\label{Ward}
\eeeq
where $\lambda$ is the helicity of the quark.
\par  We make a few observations. First is that if we look at the products with $q'$ and $p$ we can see that: $qV\left(\beta\right)=\left(q'-xp\right)V\left(\beta\right)=0$, showing that the Ward-Takahashi identity is satisfied. 
What happens is that if we multiply the diagrams and and c by $q^\mu$, the denominators proportional to 
$1/(x_4+\beta)$ and $1/(x_3-\beta-x)$ cancel out with nominators of the diagrams. We are left then with the some 
of two terms one proportional to $1/(x_1\beta s-\vec k_t^2)$ and $1/(-x_2(\beta+x)s-\vec k_t^2$ due to the diagrams a and c respectively.The sum of this terms gives the first line in Eq. \ref{Ward}, the two multipliers give the convergent contribution
  $$\sim \frac{1}{(x_1\beta s-\vec k_t^2)(-x_2(\beta+x)s-\vec k_t^2)} $$, that is cancelled by the contribution of the diagram b.
\par Note that although the diagrams a and c seem to be divergent in $\beta$ and cn not be calculated using residues, due to $\beta$ in the nominators, their sum, convoluted with q, has no dependence on $\beta$ in the nominator and is convergent and can be calculated usingthe residue analysis.
\par Second,  in the transverse products, meaning $k_tV\left(\beta\right)$ and $\kappa V\left(\beta\right)$, we find that each amplitude has  basic poles in $\beta$, but  for amplitude (b) they are always on different sides of the real line, while  for amplitudes (a) and (c) both poles are on the same side meaning they don't contribute to the amplitudes of the physical processes. Note that $\beta$ dependence in the denominators is also cancelled out, permitting the application of the residue technique.

\end{document}